\documentclass[useAMS,usenatbib]{mn2e}
\usepackage{graphicx}
\usepackage{epsfig}
\usepackage{amsmath}
\usepackage{mathrsfs}
\usepackage{amssymb}
\usepackage[english]{babel}
\usepackage{amsfonts}
\usepackage{fontenc}
\usepackage{times}
\usepackage{txfonts}
\usepackage{textcomp}
\usepackage{booktabs}
\usepackage{microtype}
\usepackage{aas_macros}
\usepackage[pdftex]{lscape}
\usepackage[usenames,dvipsnames,svgnames,table]{xcolor}

\newcommand*{\e}[1]{\times 10^{#1}}
\newcommand*{\beq}{\begin{equation}}
\newcommand*{\eeq}{\end{equation}}
\newcommand*{\Bld}[1]{\boldsymbol{#1}}
\newcommand*{\avg}[1]{\langle{#1}\rangle}
\newcommand*{\se}{\sigma_{\mathrm{e8}}}
\newcommand*{\re}{R_{\mathrm e}}
\newcommand*{\reo}{R_{\mathrm e\,0}}
\newcommand*{\Lb}{L_{\mathrm B}}
\newcommand*{\Lbs}{L_{\mathrm{B}\,\odot}}
\newcommand*{\Lv}{L_{\mathrm V}}
\newcommand*{\Lvs}{L_{\mathrm{V}\,\odot}}
\newcommand*{\Lr}{L_{\mathrm r}}
\newcommand*{\Lk}{L_{\mathrm K}}
\newcommand*{\Lks}{L_{\mathrm{K}\,\odot}}
\newcommand*{\Lx}{L_{\mathrm X}}
\newcommand*{\Tx}{T_{\mathrm X}}
\newcommand*{\Msun}{M_{\odot}}

\newcommand*{\ergs}{\mathrm{erg~s}^{-1}}
\newcommand*{\kms}{\mathrm{km~s}^{-1}}
\newcommand*{\scen}{\sigma_{\mathrm c}}
\newcommand*{\Tsn}{T_{\mathrm{SN}}}
\newcommand*{\Lsn}{L_{\mathrm{SN}}}
\newcommand*{\Tgp}{T^+_{\mathrm{g}}}
\newcommand*{\Tgm}{T^-_{\mathrm{g}}}
\newcommand*{\Lgm}{L^-_{\mathrm{g}}}
\newcommand*{\Egp}{E^+_{\mathrm{g}}}
\newcommand*{\Egm}{E^-_{\mathrm{g}}}
\newcommand*{\Tinj}{T_{\mathrm{inj}}}
\newcommand*{\gth}{\gamma_{\mathrm{th}}}
\newcommand*{\fDM}{f_{\mathrm{DM}}}
\newcommand*{\dV}{\,\mathrm{d}V}
\newcommand*{\Vmax}{V_{\mathrm{max}}}
\newcommand*{\Tsub}{T_{\mathrm{esc}}^{\mathrm{sub}}}
\newcommand*{\kb}{k_{\mathrm B}}
\newcommand*{\vphi}{v_{\varphi}}
\newcommand*{\Vrms}{V_{\mathrm{rms}}}
\newcommand*{\Vrmse}{V_{\mathrm{rms,\,e8}}}

\title[X-ray haloes and galaxy structure]
{The Effects of Galaxy Shape and Rotation on the X-ray Haloes of
Early-Type Galaxies}
 \author[S. Posacki, S. Pellegrini \& L. Ciotti]{Silvia Posacki\thanks{E-mail:
silvia.posacki@unibo.it}, Silvia Pellegrini \& Luca Ciotti
\\Department of Physics and Astronomy, University of Bologna, viale Berti Pichat
6/2, 40127 Bologna, Italy}

\date{Accepted 2013 May 17. Received 2013 May 16; in original form 2013 March 28}

\pagerange{\pageref{firstpage}--\pageref{lastpage}} \pubyear{2013}

\begin{document}
\maketitle
\label{firstpage}

\begin{abstract}
We present a detailed diagnostic study of the observed temperatures of the hot
X-ray coronae of early-type galaxies. 
Extending the investigation carried out in \citet{Pel} with spherical models,
we focus on the dependence of the energy budget and temperature of 
the hot gas on the galaxy structure and internal stellar kinematics. By 
solving the Jeans equations we construct realistic axisymmetric three-component
galaxy models (stars, dark matter halo, central black hole) with different
degrees of flattening and rotational support. The kinematical fields are
projected along different lines of sight, and the aperture velocity dispersion 
is computed within a fraction of the circularized effective radius. 
The model parameters are chosen so that the models resemble real ETGs and
lie on the Faber--Jackson and Size--Luminosity relations. 
For these models we compute $T_*$ (the stellar heating contribution to the gas injection temperature) and $\Tgm$
(the temperature equivalent of the energy required for the gas escape). In particular,
different degrees of thermalisation of the ordered rotational field of the galaxy are considered.
We find that $T_*$ and $\Tgm$ can vary only mildly due to a pure change of shape.
Galaxy rotation instead, when not thermalised, can lead to a large decrease of $T_*$;
this effect can be larger in flatter galaxies that can be more rotationally supported.
Recent temperature measurements $\Tx$, obtained with \textit{Chandra}, are larger than,
but close to, the $T_*$ values of the models, and show a possible trend for a lower $\Tx$ in flatter and
more rotationally supported galaxies; this trend can be explained by the lack of thermalisation
of the whole stellar kinetic energy. Flat and rotating galaxies also show lower $\Lx$ values, and
then a lower gas content, but this is unlikely to be due to the small variation of $\Tgm$ found here
for them.

\end{abstract}


\section{Introduction} \label{sec:intro}

High-quality X-ray observations of early-type galaxies (ETGs)
performed with the \textit{Chandra} X-ray
Observatory have produced a large body of data for the study of the hot gas
haloes in these galaxies with unprecendented detail. In particular,
the nuclear and the stellar (resolved and unresolved) contributions to
the total X-ray emission could be subtracted, obtaining more accurate
properties of the hot interstellar medium (ISM) than ever before. From
a homogeneous and thorough X-ray analysis for the pure gaseous
component, samples of ETGs have been built with an improved measurement of the X-ray average
temperature $\Tx$ and luminosity $\Lx$ for the gas only (e.g., \citealt[hereafter BKF]{Boro}).
This X-ray information can now be
compared with that provided by other fundamental properties of ETGs,
such as their total optical luminosity ($\Lb$ or $\Lk$ in the $B$ or
$K$ band), central stellar velocity dispersion ($\scen$), galaxy
rotation and shape, to explore and revisit the possible relations
betweeen the main galactic properties and those of the X-ray gas. For
example, in the $\Lx-\Lk$ correlation, the previously known large
variation of up to two orders of magnitude in $\Lx$ at the same $\Lk$
has been even extended, due to the inclusion of hot-gas poor ETGs
in a larger fraction than previously possible (with $\Lx$ extending down to $\sim
10^{38}$ $\ergs$; BKF). This variation is related to the ISM 
evolution over cosmological time-scales, during which stellar
mass losses and explosions of Type Ia supernovae (SNIa) provide
gas and gas heating, respectively, to the hot haloes (possibly in
conjunction with feedback from accretion on to the central
supermassive black hole, SMBH). Modulo environmental effects like galaxy
interactions or tidal stripping, the hot gas content and
temperature fundamentally depend on the energy budget of the hot ISM, that
in turn depends on the particular host galaxy structure and internal
kinematics (e.g., \citealt[hereafter CDPR]{CDPR}; \citealt[hereafter
P11]{Pel}). For
example, the luminous and dark matter content and
distribution determine the potential well shape and depth, and so the
binding energy of the gas and its dynamical state (e.g., \citealt[hereafter CP96]{CP96}).
Indeed, one of the discoveries that followed the analysis of first X-ray data of ETGs
was the sensitivity of the hot gas content to major galaxy properties as the shape of the mass
distribution, and the mean rotation velocity of the stars (see
\citealt{Pelbook} for a review). The investigation of the origin of
this sensitivity is the goal of the present paper.

A relation between the hot gas retention capability and the intrinsic galactic shape
became apparent already in the X-ray sample of ETGs built from \textit{Einstein}
observations: on average, at any fixed $\Lb$, rounder
systems had larger total $\Lx$ and $\Lx/\Lb$, a measure of the galactic hot
gas content, than flatter ETGs and S0
galaxies \citep{Eskridge95}. Moreover, galaxies with axial ratio
close to unity spanned the full range of $\Lx$, while flat systems
had $\Lx\lesssim 10^{41}\ergs$. This result was not produced by flat
galaxies having a lower $\Lb$, with respect to round ETGs,
since it held even in the range of $\Lb$ where the two shapes coexist
\citep{Pel99}. The relationship between $\Lx$ and shape was
reconsidered, confirming the above trends, for the \textit{ROSAT} PSPC sample
\citep{Pelbook}, and for the \textit{Chandra} sample \citep{LiWang}.
Therefore, there seems to be an empirical dependence of the hot gas
content on the galactic shape, and it was suggested that a flatter
shape by itself may be linked to a less negative binding energy for the gas (CP96).
However, since flatter systems also possess a higher rotational
support on average (e.g., \citealt{BinTre}), also the influence of galactic rotation
on the hot gas was called into question.
For example, in rotationally supported ETGs the gas may be less bound, compared to the ISM in
non-rotating ETGs, leading rotating ETGs to be more prone to host outflowing regions.
For these reasons, the effects on $\Lx$ of both galactic
shape and rotation were studied for a sample of 52 ETGs with known
maximum rotational velocity of the stars $\Vmax$, and so with a measure of $\Vmax/\scen$,
an indicator of the importance of rotation \citep{Pel97}. It was found that
$\Lx/\Lb$ can be high only for $\Vmax/\scen <~ 0.4$, and is limited
to low values for $\Vmax/\scen>0.4$. This trend was not produced by being the ETGs
with high $\Vmax/\scen$ confined to low $\Lb$. 
\citet{Sarzi10} investigated again the relationship between
X-ray emission (from \textit{Einstein} and \textit{ROSAT} data) and rotational
properties for the ETGs of the \textit{SAURON} sample, confirming that
slowly rotating galaxies can exhibit much larger luminosities than
fast-rotating ones.

Renewed interest in the subject has come recently after the higher
quality \textit{Chandra} measurements of $\Lx$ and $\Tx$
have become available. In an investigation using
\textit{Chandra} and \textit{ROSAT} data for the ATLAS$^{\mathrm{3D}}$ sample,
\citet[hereafter S13]{Sarzi13} found that slow rotators generally have the largest $\Lx$ and
$\Lx/\Lk$ values, and $\Tx$ values consistent just with the
thermalisation of the stellar kinetic energy, estimated from $\sigma _{\mathrm e}$
(the stellar velocity dispersion averaged within the optical effective radius $\re$). Fast
rotators, instead, have generally lower $\Lx$ and $\Lx/\Lk$ values,
and the more so the larger their degree of rotational support; the
$\Tx$ values of fast rotators keep below 0.4 keV and do not scale with $\sigma _{\mathrm e}$
(see also BKF). Considering that fast rotators are likely to be intrinsically flatter than slow
rotators, and that the few slow rotators with low $\Lx$ are also
relatively flat, S13 supported the hypothesis
whereby flatter galaxies have a harder time in retaining their hot gas
(CP96). 
To explain why fast rotators seem confined to lower $\Tx$ than slow rotators,
they suggest that the kinetic energy associated with the stellar ordered motions
may be thermalised less efficiently.

In order to help clarify what is the expected variation of hot gas content
and temperature, originating in a variation of shape and internal kinematics
(and in its degree of thermalisation) of the host galaxy,
we embarked on an investigation based on the numerical building of state-of-the-art
galaxy models, and of the associated temperature and energy budget for the gas.
In Sect.~\ref{sec:temp} we define a set of mean temperatures for the models,
some of which already introduced in P11. In Sect.~\ref{sec:mods} we describe the different profiles
adopted for the mass components of the galaxy models, the scaling laws considered to
constrain the models to resemble real galaxies, the observable
properties of the models in the optical band, and the procedure to obtain flat models.
Our main results are presented in Sect.~\ref{sec:res}, together with a comparison with the observed X-ray
properties of ETGs. Finally Sect.~\ref{sec:conclu} presents our main conclusions.
Appendix~\ref{sec:flueqs} recalls the fluid equations
in the presence of source terms, and Appendix~\ref{sec:code} summarizes the main
procedural steps of the code built on purpose for the construction of
the models.

\section[]{The temperatures} \label{sec:temp}
Here we introduce a set of gas mass-weighted temperatures, equivalent to the
injection and binding energies of the hot gas in ETGs.

\subsection{The injection temperature}

In the typically evolved stellar population of ETGs, the main processes
responsible for the injection of gas mass, momentum and energy in the ISM are stellar
winds from red/asymptotic giant branch stars, and Type Ia supernova explosions (SNIa), 
the only ones observed in an old stellar population (e.g. \citealt{Capp}).
The wind material outflowing from stars leaves the stellar surface with
low temperatures and low average velocities ($\sim$ few 10 $\kms$; \citealt{Parriott}), so that
all its energy essentially comes from the stellar motion inside the galaxy.
SNIa's explosions, instead, provide mass to the ISM through their very high velocity ejecta ($\sim$ few $10^{4}$ $\kms$).
Thus, this material provides mass and heat to the hot haloes, via thermalisation of its energy
through shocks with the ambient medium or with other ejecta, heating up to X-ray
emitting temperatures.

The injection energy per unit mass due to both heating processes is 
$e_{\mathrm{inj}}\equiv3k_{\mathrm B}\Tinj/(2\mu m_{\mathrm P}$), where
$k_{\mathrm B}$ is the Boltzmann constant, $\mu=0.62$ is the mean molecular weight for
solar abundance, $m_{\mathrm P}$ is the proton mass, and $\Tinj$ is defined as 
\beq
\Tinj\equiv \dfrac{\dot{M}_*T_{*} + \dot{M}_{\mathrm{SN}}\Tsn}{\dot{M}}.
\eeq
Here $T_*$ and $\Tsn$ are the injecta temperatures resulting from
the thermalisation of their interactions with the ISM through stellar winds
and SNIa's respectively (see below).
$\dot{M}$ is the total mass-loss rate for the entire galaxy, given by the sum of the
stellar mass-loss rate $\dot{M}_*$ and of the rate of mass
loss via SNIa events $\dot{M}_{\mathrm{SN}}$ 
($\dot{M}=\dot{M}_*+\dot{M}_{\mathrm{SN}}$).
The time evolution of the stellar mass-loss rate $\dot{M}_*$ can be calculated using
single-burst stellar population synthesis models for different initial mass
functions (IMFs) and metallicities (e.g., \citealt{Mara}).
For example, at an age of 12 Gyr, $\dot{M}_* (\Msun$ yr$^{-1})\approx 2\e{-11}\, \Lb(\Lbs)$
 for the Salpeter or Kroupa IMF (e.g., \citealt{Pelbook}).
$\dot{M}_{\mathrm{SN}}$ is instead given by $\dot{M}_{\mathrm{SN}}=M_{\mathrm{SN}}R_{\mathrm{SN}}$, where
$M_{\mathrm{SN}}=1.4\,\Msun$ is the mass ejected by one SNIa event and 
$R_{\mathrm{SN}}$ is the explosion rate. For local ETGs it is
$R_{\mathrm{SN}}=0.16(H_0/70)^2 \e{-12}\,\Lb(\Lbs)\,\mathrm{yr}^{-1}$, 
where $H_0$ is the Hubble constant in units of
$\kms\mathrm{Mpc}^{-1}$ \citep{Capp}. More recent measurements of
the observed rates of supernovae in the local universe \citep{Li} give a SNIa
rate in ETGs consistent with that of \cite{Capp}. For this rate, and
$H_0=$70 $\kms\textrm{Mpc}^{-1}$, one obtains
$\dot{M}_{\mathrm{SN}}=2.2 \e{-13}\,\Lb(\Lbs)$ $\Msun\,\mathrm{yr}^{-1}$,
which is $\sim 80$ times smaller than the $\dot{M}_*$ above for an age
of 12 Gyr. Thus the main source of mass is provided by $\dot{M}_*$, and
approximating $\dot{M}\simeq\dot{M}_*$, we have $\Tinj\simeq
T_*+(\dot{M}_{\mathrm{SN}}/\dot{M}_*)\Tsn$.

Neglecting the internal energy and the stellar wind velocity relative to the star,
$T_*$ is the sum of two contributions, deriving from the random and
the ordered stellar motions. In axisymmetric model
galaxies as built here, the stellar component of the galaxy is allowed
to have a rotational support, and the latter can be
converted into heating of the injected gas in a variable amount. The extent of
the contribution of rotational motions is not known a priori, since it depends on both the
importance of the stellar ordered motions and the dynamical status of the
surrounding gas already in situ (see Appendix~\ref{sec:flueqs}; see also \citealt{Derc,ProcAnd}). 
Given the complexity of the problem, hydrodynamical
simulations are needed to properly calculate this heating term, but we
can still obtain a simple estimate of it by making reasonable assumptions.
We define the equivalent temperature of stellar motions $T_*$ as
\beq
T_*= T_{\sigma}+\gth T_{\mathrm{rot}}
\label{eq:tstar}
\eeq
where
\beq
T_{\sigma}=\dfrac{\mu
m_{\mathrm P}}{3k_{\mathrm B} M_*} \int\rho_*\mathrm{Tr}(\Bld{\sigma}^2)\dV 
\label{eq:tsig}
\eeq
is the contribution of stellar random motions, 
\beq
T_{\mathrm{rot}}=\dfrac{\mu m_{\mathrm P}}{3k_{\mathrm B} M_*}\int\rho_*\overline{\vphi}^2\dV 
\label{eq:trot}
\eeq
is the one due to the stellar streaming motions, and $\gth$ is a
parameter that regulates the degree of thermalisation of the ordered
stellar motions. $M_*$ is the stellar mass of the galaxy,
$\Bld{\sigma}^2$ is the velocity dispersion tensor, and $\overline{\vphi}$ is
the azimuthal component and the only non-zero component of the
streaming velocity $\Bld{v}=\overline{\vphi}\mathbf{e}_{\varphi}$ (see Appendix~\ref{sec:code}). 
In Eqs.~(\ref{eq:tsig}) and (\ref{eq:trot}), as in the remainder of the paper,
we assume that the gas is shed by stars with a spatial dependence that
follows that of the stellar distribution $\rho_*$, so that the density
profile of the gas injected per unit time is proportional to $\rho_*$, i.e. it is
$\mathscr{M}=\dot{M}\rho_*/M_*$ in Eqs.~(\ref{eq:1fe}) $-$ (\ref{eq:3fe}) of
Appendix~\ref{sec:flueqs}. 

The parameter $\gth$ is defined as
\beq
\gth =\dfrac{\mu m_{\mathrm P}}{T_{\mathrm{rot}}3k_{\mathrm B} M_*} \int\rho_*\Arrowvert
\Bld{u}- \Bld{v} \Arrowvert ^2 \dV, 
\label{eq:gth}
\eeq
where $\Bld{u}$ is the velocity of the pre-existing gas (see Appendix~\ref{sec:flueqs}).
A simple estimate for $\gth$ is obtained when the gas velocity is proportional to
$\Bld{v}$, i.e., $\Bld{u}=\alpha\Bld{v}$, where $\alpha$ is some constant.
In this special case, from Eqs.~(\ref{eq:trot}) and (\ref{eq:gth}) it follows
that $\gth=(\alpha -1)^2$. When $\alpha=1$, gas and stars rotate with
the same velocity and no ordered stellar kinetic energy is thermalised,
whereas for $\alpha=0$ the gas is at rest and all the kinetic energy of the stars,
including the whole of the rotational motions, is thermalised. 
Clearly both cases are quite extreme and unlikely, and plausibly the pre-existing
gas will have a rotational velocity ranging from zero to the streaming velocity of stars, i.e.,
$0\leqslant\alpha\leqslant1$ and then $1\geqslant\gth\geqslant0$ (the case of a constant $\alpha>1$,
where the pre-existing gas is everywhere rotating faster than the
newly injected gas is not considered). Note that, contrary to $\gth T_{\mathrm{rot}}$,
$T_{\sigma}$ is in principle exact, and can be computed a priori.

The internal plus kinetic energy of the ejecta, released during a SNIa event, is of the order of
$E_{\mathrm{SN}}\simeq 10^{51}$ erg. Depending on the conditions of the
environment in which the explosion occurs, the radiative losses from
the expanding supernova remnant may be important, and so the amount of
energy transferred to the ISM through shock heating is a fraction
$\eta$ of $E_{\mathrm{SN}}$. Realistic values of $\eta$ for the hot
and diluted ISM of ETGs are around $0.85$ (e.g., \citealt{Tang},
\citealt{Thornton}), thus
\beq
\Tsn=\dfrac{2\mu m_{\mathrm P}}{3k_{\mathrm B}}\dfrac{\eta E_{\mathrm{SN}}}{M_{\mathrm{SN}}},
\label{eq:tsn}
\eeq
and, substituting the above expressions for $\dot{M}_{\mathrm{SN}}$ and $\dot{M}_*$, 
we obtain the average injection temperature
\beq
\Tinj=T_*+1.7\dfrac{\eta}{0.85} \e{7}\,\textrm{K}.
\label{eq:tinj}
\eeq

A possible additional source of heating for the gas could be provided
by a central SMBH. Through its gravitational influence, it is
responsible for the increase of the stellar motions within its radius
of influence (of the order of a few tens of parsecs; e.g.,
\cite{Pelbook}). We consider this effect here, while we neglect
possible effects as radiative or mechanical feedback.

\subsection{The temperatures related to the potential well}\label{sec:gravtemp}

The gas ejected by stars can be also heated `gravitationally' by falling
into the galactic potential well to the detriment of its potential energy, and
by the associated adiabatic compression. When stellar mass losses accumulate, the gas
density can reach high values and the cooling time can become smaller than
the galactic age; if the radiative losses increase considerably,
the gravitational force overwhelms the pressure gradient and eventually the gas
starts inflowing toward the centre of the galaxy. 
Thus we can define a temperature
\beq
\Tgp= \dfrac{2\mu m_{\mathrm P}\Egp}{3k_{\mathrm B}}=\dfrac{2\mu m_{\mathrm P}}{3k_{\mathrm B}M_*}
\int \rho_*\,(\Phi-\Phi_0) \dV
\label{eq:egr+}
\eeq
where $\Egp$ is the average change in gravitational energy per unit mass of the gas
flowing in through the galactic potential $\Phi(\mathbf{x})$
down to the galactic centre, and $\Phi_0=\Phi(0)$. Note that $\Egp>0$, having
assumed as usual that $\Phi(\mathbf{x})<0$.
However, as discussed in P11, most of $\Egp$ may be radiated away, and there
are conditions under which Eq.~(\ref{eq:egr+}) does not apply. 
Therefore, given these uncertainties, we consider $\Tgp$ just as a
reference value, and keep in mind that the temperature achievable from infall 
can be much lower than that given by Eq.~(\ref{eq:egr+}).

By analogy with $\Tgp$, we can define a temperature
\beq
\Tgm= \dfrac{2\mu m_{\mathrm P}\Egm}{3k_{\mathrm B}}=-\dfrac{2\mu m_{\mathrm P}}{3k_{\mathrm B}M_*}\int \rho_*\Phi\dV,
\label{eq:tgr-}
\eeq
where $\Egm$ is the average energy necessary to extract a unit of gas mass
from the galaxy,
with the assumption that $\Phi(\infty)=0$.
If the gas rotates, Eq.~(\ref{eq:tgr-}) must be modified since, thanks to the
centrifugal support, the gas is less bound. Assuming again that $\Bld{u}=\alpha\Bld{v}$, then
\beq
\Egm(\alpha)= -\dfrac{1}{M_*}\int
\rho_*\left(\Phi +\dfrac{\alpha^2}{2}
\overline{\vphi}^2\right)\dV,
\label{eq:egr-rot}
\eeq
so that $\Tgm(\alpha)=\Tgm-\alpha^2 T_{\mathrm{rot}}$.
Note that, for a given $T_{\mathrm{rot}}$, the smallest is $\gth$ (the largest is $\alpha$),
the smallest is $T_*$ (the gas is less heated),
but also the lower is $\Tgm$ (the gas is less bound).

When the galaxy mass distribution has a potential that diverges at
small and/or large radii, as for the singular isothermal sphere, we
assume the gas has been extracted from the galaxy when it has reached a distance of
15 $\re$ from the galactic centre, so that $\Egm$ and $\Egp$ do not correspond exactly to
Eqs.~(\ref{eq:egr-rot}) and (\ref{eq:egr+}). 
Finally, if energy losses due to cooling are present, the gas would
need more than $\Egm$ to escape, but these losses are negligible for outflows that
typically have a low density. 

In case of gas escape, we can introduce another mass-weighted temperature
by considering the enthalpy per unit mass of a perfect gas
$h=\gamma k_{\mathrm B}T/[\mu m_{\mathrm P}(\gamma-1)]=c_s^2/(\gamma-1)$,
where $\gamma$ is the ratio of the specific heats and $c_s$ is the sound speed.
Building on the Bernoulli theorem, we can
derive a fiducial upper limit to the temperature of outflowing gas. For a fixed galactic potential, the energy
of the escaping gas can be divided between kinetic and thermal energy with
different combinations.
In the extreme case in which the gas reaches infinity with a null 
velocity and enthalpy, and it is injected with a (subsonic) velocity $\Bld{u}=\alpha\Bld{v}$, 
from the Bernoulli equation $h(\mathbf{x})+v^2(\mathbf{x})/2+\Phi(\mathbf{x})=0$,
we derive a characteristic gas-mass averaged escape temperature
\beq
\Tsub =-\dfrac{2\mu m_{\mathrm P}}{5k_{\mathrm B}M_*}\int \rho_*
\left(\Phi +\dfrac{\alpha^2}{2}
\overline{\vphi}^2\right)
\dV=\dfrac{3}{5} \Tgm(\alpha),
\label{eq:tsub}
\eeq
for a monoatomic gas (see P11 for more details). In the opposite case of an important kinetic energy of the flow, 
the gas temperature will be lower than $\Tsub$.

In summary, $\Tgm$ is a temperature equivalent to the
energy required to extract the gas, while $\Tsub$ is close to the temperature we
expect to observe for outflowing gas. For inflowing gas, we expect to
observe a temperature much lower than $\Tgp$, since more than
$\sim 0.5 \Egp$ is radiated away or goes into kinetic
energy of the gas, or because of condensations in the gas (e.g., \citealt{Sarazin89}). For reference, for
realistic spherical models, $\Egp \sim 2
\Egm$, thus the temperature of the inflowing gas should be lower than $\Tgm$ (P11).

Finally, we mention about the relation between observed
temperatures $\Tx$ and the average mass-weighted temperatures of
this Section. The latter are derived under the
assumption that the gas density $\rho_{\mathrm{gas}}$ follows that of the stars,
which is appropriate for the continuously injected gas (e.g., for $T_*$ and
$\Tinj$), while the bulk of the hot ISM may have a different distribution; thus, mass-weighted
$\Tgm$ and $\Tsub$ referring to the whole hot gas content of an ETG may be different from those given by
Eqs.~(\ref{eq:tgr-}) and~(\ref{eq:tsub}). In general, the $\rho_{\mathrm{gas}}$
profile is shallower than that of $\rho_*$ (e.g., \citealt{Sarazin88,Fabbiano89}), and then
the gas mass-weighted $\Tgp$ would be larger than derived with
Eq.~(\ref{eq:egr+}), and the mass-weighted $\Tgm$ or $\Tsub$ would be lower than derived using
Eqs.~(\ref{eq:tgr-}) and~(\ref{eq:tsub}). For steady winds, instead,
when the gas is continuously injected by stars and expelled from the
galaxy, the $\rho_{\mathrm{gas}} \propto \rho_*$ assumption is a good approximation.

Another point is that the $\Tx$ values are emission-weighted
averages, and will coincide with mass-weighted averages only if the
entire ISM has one temperature value (e.g., \citealt{CioPel08}; \citealt{Kim12}).
A single $\Tx$ value measured from the spectrum of the integrated emission
will tend to be closer to the temperature of the
densest region, in general the central one, thus it will be closer to the
central temperature than the mass-weighted one. The temperature
profiles observed with $Chandra$ tend to be quite flat, except for
cases where they increase outside of $\sim 0.5 \re$
(generally in ETGs with the largest $\Tx$), and for cases of negative temperature
gradients (in ETGs with the lowest $\Tx$; \citealt{Diehl,Nagino}).
Therefore, the largest $\Tx$ may be lower than mass-weighted
averages, and the lowest $\Tx$ may be larger than them. In conclusion, the comparison of $\Tx$ and the gas content
with the gas temperature and binding energy introduced in this Section (as $T_*$ and $\Egm$)
represents the easiest approach for a general, systematic
investigation involving a wide set of galaxy models,
but the warnings above should be kept in mind. Note, however, that
the conclusions below remain valid when taking into account the above considerations.

\section[]{The models} \label{sec:mods}

The galaxy models used for the energetic estimates include three components: a stellar distribution, a dark matter 
(DM) halo, and a central SMBH. The stellar component is axisymmetric and can have
different degrees of flattening, while for simplicity the DM halo is 
spherical. The SMBH is a central mass concentration with mass $M_{\mathrm{BH}}=10^{-3} M_*$,
following the \cite{Mag} relation. Its effects are minor, but it is considered
for completeness. For these models,
the Jeans equations are solved under the standard assumption of a two-integral
phase space distribution function (see Appendix~\ref{sec:code}), so that, besides
random motions, stars can have ordered motions only in the azimuthal direction.
The decomposition of the azimuthal motions in velocity dispersion and streaming
velocity is performed via the $k$-decomposition introduced by
\cite{Satoh}; thus, the amount of rotational support is varied simply through
the parameter $k$. With the adoption of the mass profiles detailed below 
for the stars and the DM, we built galaxy models that reproduce with a good
level of accuracy the typical properties of the majority of ETGs.
The models are then projected along two extreme lines of sight (corresponding to the face and edge-on views) 
and forced to resemble real galaxies as described in Section~\ref{sec:obs}.

\subsection[]{Stellar distribution} \label{sec:mods_star}

The stellar distribution is described by the \cite{devauc} law, by using the
deprojection of \cite{MelMat} generalized for ellipsoidal axisymmetric distributions
\beq
\rho_*(R,z)=\rho_0\xi^{-0.855}\exp(-\xi^{1/4}),
\label{eq:rho_*}
\eeq
with
\beq
\rho_0=\dfrac{M_*b^{12}}{16\pi q\reo^3\Gamma(8.58)},\quad \xi=\dfrac{b^4}{\reo}\sqrt{R^2+\dfrac{z^2}{q^2}},
\label{eq:rho_*2}
\eeq
where $(R,\varphi,z)$ are the cylindrical coordinates and $b\simeq7.66925$.
The flattening is controlled by the parameter $q\leqslant 1$, so that the minor
axis is aligned with the $z$ axis.
$\reo$ is the projected half mass radius (effective radius) when the galaxy is seen face-on;
for an edge-on view, the circularized effective radius is $\re=\reo\sqrt{q}$ (Sect.~\ref{sec:flat} and Appendix~\ref{sec:code}).
We assume a constant stellar mass-to-light ratio $\Upsilon_*$ all over the galaxy,
so that $M_*$ is directly proportional to the luminosity $L$.
Note that Eq.~(\ref{eq:rho_*2}) guarantees that the total stellar mass (luminosity) of the model
is independent of the choice of $q$ and $\reo$.

\subsection[]{Dark matter halo}\label{sec:mods_halo}

Given the uncertainties affecting our knowledge of the density
profile of DM haloes, we explored four families of DM profiles.
The first one is the scale-free singular isothermal sphere (SIS)
\beq
\rho_{\mathrm{h}}(r)=\dfrac{v_{\mathrm c}^2}{4\pi Gr^2}, \qquad\qquad \Phi_{\mathrm{h}}(r)=v_{\mathrm c}^2\ln r,
\label{eq:SIS}
\eeq
where $v_{\mathrm c}$ is the halo circular velocity. The gravitational potential
of this profile diverges at small and large radii, thus the potential is truncated at 
a distance of 15 $\re$ to obtain a finite $\Tgm$.

A number of recent works are reconsidering the \cite{Ein} profile as 
appropriate to model DM haloes (e.g. \citealt{Navarro04,Merr,Gao08,Navarro10}).
The density distribution of this profile is the
three-dimensional analogue of the S\'{e}rsic law, widely used to fit the surface
brightness profiles of ETGs. The density is described by
\beq
\rho_{\mathrm{h}}(r)=\rho_{\mathrm c}\exp(d_n-x),
\label{eq:Einasto}
\eeq
where $\rho_{\mathrm c}$ is the density at the volume half-mass radius $r_{\mathrm{h}}$,
$x\equiv d_n(r/r_{\mathrm{h}})^{1/n}$, $n$ is a free parameter, and $d_n$ is well approximated by the relation
\beq
d_n\simeq 3n-\frac{1}{3}+\frac{8}{1215~n},
\eeq
\citep{RetMon}. Finally the gravitational potential is
\beq
\Phi_{\mathrm{h}}(x)=-\dfrac{GM_{\mathrm{h}}}{r}\left[1-\dfrac{\Gamma(3n,x)}{\Gamma(3n)}+\dfrac{
x^n\Gamma(2n,x)}{\Gamma(3n)} \right].
\eeq

The third family is based on the \cite{Hern} profile
\beq
\rho_{\mathrm{h}}(r)=\dfrac{M_{\mathrm{h}} r_{\mathrm{h}}}{2\pi r(r+r_{\mathrm{h}})^3}, \qquad
\Phi_{\mathrm{h}}(r)=-\dfrac{GM_{\mathrm{h}}}{r+r_{\mathrm{h}}},
\eeq
where $M_{\mathrm{h}}$ and $r_{\mathrm{h}}$ are the halo total mass and scale radius, respectively.

Lastly, we used also the NFW profile \citep{Nfw}
\beq
\rho_{\mathrm{h}}(r)=\dfrac{\rho_{\mathrm{crit}}~\delta_{\mathrm c}r_{\mathrm{h}}}{r\left(1+r/r_{\mathrm{h}}\right)^2},
\eeq
where $\rho_{\mathrm{crit}}=3H^2/8\pi G$ is the critical density for closure.
The total mass diverges, so it is common use to
identify the characteristic mass of the model $M_{\mathrm{h}}$ with the mass enclosed
within $r_{200}$, defined as the radius of a sphere of mean interior density
200 $\rho_{\mathrm{crit}}$. Then, from the definition of $r_{200}$, the
concentration $c\equiv r_{200}/r_{\mathrm{h}}$ and the coefficient $\delta_{\mathrm c}$ are linked as
\beq
\delta_{\mathrm c}=\frac{200}{3}\dfrac{c^3}{\ln(1+c)-c/(1+c)}.
\eeq
The gravitational potential of the NFW profile is
\beq
\Phi_{\mathrm{h}}(r)=-4\pi G\rho_{\mathrm{crit}}~\delta_{\mathrm c}r_{\mathrm{h}}^3~\dfrac{\ln (1+r/r_{\mathrm{h}})}{r}.
\eeq

\subsection[]{Linking the models to real ETGs} \label{sec:obs}
One of the most delicate steps of the present
study is to have a sample of galaxy models, characterised by various degrees of flattening and
rotational support, that closely resemble real ETGs, at least in a statistical sense. This
is accomplished by flattening spherical models, that we call `progenitors'.

In fact, the process of flattening a galaxy model is not trivial, and it is highly degenerate,
as illustrated by the exploratory work of CP96, where the full parameter space of two-component
Miyamoto-Nagai models was explored. Here, we begin with a generic spherical galaxy model, and
we impose that its effective radius $\re$ and aperture luminosity-weighted
velocity dispersion within $\re/8$, $\se$, satisfy the most important observed scaling laws (SLs)
of ETGs, the Faber--Jackson and the Size--Luminosity
relations. In particular, we use the Faber--Jackson and the Size--Luminosity relations derived in the $r$ band for
$\approx80\,000$ ETGs drawn from Data Release 4 (DR4) of the Sloan
Digital Sky Survey (SDSS; \citealt{Desroches}). These relations are
quadratic best-fitting curves, with a slope varying with luminosity $\Lr$:
\beq \log\se=-1.79 +0.674\log \Lr-0.0234\log^2 \Lr,
\label{eq:scalelaws0}
\eeq 
\beq
\log \re=1.50-0.802\log \Lr+0.0805\log^2 \Lr,
\label{eq:scalelaws}
\eeq
where $\se$ and $\re$ are in units of $\kms$ and $\textrm{kpc}$
respectively, and $\Lr$ is calibrated to the AB system
\citep{Desroches}.

In practice, we fix a value for $\se$ in the range 150 $\kms\lesssim\se\lesssim300$ $\kms$,
and then we derive $\Lr$ and $\re$ from Eqs.~(\ref{eq:scalelaws0}) and~(\ref{eq:scalelaws}).
After conversion of $\Lr$ to the $V$-band\footnote{The $V$-band luminosity $\Lv$ is computed using 
the standard transformation equations between SDSS magnitudes and other systems 
(http://www.sdss3.org/dr9/algorithms/sdssUBVRITransform.php),
also assuming $B-V=0.9$ as appropriate for ETGs \citep{Donas}.} ($\Lv$), we derive $M_*$ 
adopting a (luminosity dependent) $V$-band mass-to-light ratio $\Upsilon_*$ appropriate 
for a 12 Gyr old stellar population with a Kroupa IMF \citep{Mara}.
Following empirical evidences \citep{Bender,Capp06}, we assume that 
$\Upsilon_*\propto \Lv ^{0.26}$, obtaining $3.3\lesssim\Upsilon_*\lesssim4.7$.
With this choice, the models need a DM halo to reproduce the assigned $\se$.
We consider the four different families of (spherical) DM haloes in Sect.~\ref{sec:mods_halo},
whose parameters are fixed to reproduce
the assigned $\se$. The simplest family is that with the SIS halo in Eq.~(\ref{eq:SIS}), where we
fix $v_{\mathrm c}$ so that the progenitor has the given $\sigma_{\mathrm e8}$.
For the Einasto DM haloes, we fix $n=6$ and $r_{\mathrm h}\simeq7\re$ in Eq.~(\ref{eq:Einasto}),
in order to obtain $r_{\mathrm h}$ values in the accepted range for ETGs
(see, e.g., \citealt{Merr,Navarro10}), and to keep low the DM fraction $\fDM$ in the central
regions of the model (see below). $M_{\mathrm h}$, the only remaining free parameter,
is then determined by the chosen $\se$. This procedure gives $M_{\mathrm h}$ values that are $\simeq20$
percent larger than in the SIS case, due to the shallower density slope of the Einasto DM halo
at small radii, which translates into a weaker effect on the stellar random motions,
and then into a larger DM amount required to raise the central stellar velocity dispersion
profile up to the chosen $\se$.
Also for the Hernquist and NFW families we choose $r_{\mathrm h}\simeq7\re$, and 
$M_{\mathrm h}$ is fixed to reproduce the assigned $\se$. For the Hernquist family, this
request results in $1.8\e{12}\Msun\lesssim M_{\mathrm h}\lesssim 4 \e{12}\Msun$, while for the NFW haloes
we find $12\lesssim c\lesssim 25$ \citep{BinTre,Napolitano}, corresponding to $10^{14}\Msun\gtrsim 
M_{\mathrm h}\gtrsim 7.2\e{12}\Msun$.
For all models, the resulting $M_{\mathrm h}/M_*$ ratios agree with those
given by cosmological simulations and galaxy mass functions \citep{Narayanan}.
A summary of the properties of some spherical progenitors, for SIS and Einasto DM haloes, is given in Table~\ref{tab:params}.
\begin{table*}
\caption{Fundamental galaxy parameters for the progenitors.}
\begin{tabular}{ccccccccccc}
\toprule
$\se$   &$\Lv$          &$\Lk$          &$\re$&  $M_*$         &$\Upsilon_*$&$v_{\mathrm c}$&$M_{\mathrm h}(15\re)$ &$\fDM $ &$M_{\mathrm h}$ &$\fDM $  \\
$(\kms)$&$(10^{11}\Lvs)$ &$(10^{11}\Lks)$ &(kpc)&$(10^{11}\Msun)$ &$(\Msun \Lvs^{-1})$    &$(\kms)$       &$(10^{11}\Msun)$        &        &$(10^{11}\Msun)$ &         \\
(1)     & (2)           & (3)           & (4) & (5)            & (6)                   & (7)           & (8)                   & (9)    & (10)           & (11)    \\
\midrule                                                                                                                                               
300     & 1.66          & 6.64          &11.79&      7.80      &   4.7                 &       237.3   &    26.57              &   0.32 &    31.53       &   0.49   \\
250     & 0.78          & 3.12          & 7.04&      3.35      &   4.3                 &       189.4   &    10.10              &   0.30 &    11.99       &   0.46   \\
200     & 0.33          & 1.32          & 4.09&      1.25      &   3.8                 &       151.9   &    3.78               &   0.30 &    4.48        &   0.46   \\
150     & 0.12          & 0.47          & 2.29&      0.39      &   3.3                 &       113.0   &    1.17               &   0.29 &    1.39        &   0.45   \\
\bottomrule
\end{tabular}
\flushleft
Notes: (1) Stellar velocity dispersion, as the luminosity-weighted average within an aperture of radius $\re/8$.
(2) and (3): luminosities in the $V$ band (derived as described in Sect.~\ref{sec:obs}) and $K$ band, from $\Lk=4\Lv$
as appropriate for a 12 Gyr old stellar population with a Kroupa IMF and solar metallicity \citep{Mara}.
(4) Effective radius. (5) Stellar mass. (6) $V$ band stellar mass-to-light ratio.
$(7)-(9)$ Circular velocity, dark matter mass within a sphere of radius 15 $\re$, and dark matter fraction
within $\re$, for the SIS halo. $(10)-(11)$ Dark matter mass and dark matter fraction within $\re$ for the Einasto halo.
\label{tab:params}
\end{table*}

An important quantity characterizing the models is the effective DM
fraction, defined as the ratio of the DM mass to the total mass
contained within a sphere of radius $\re$,
$\fDM=M_{\mathrm{h}}(\re)/M_{\mathrm{tot}}(\re)$. We compute $\fDM$ a
posteriori, to check that it agrees with the values found for well
studied ETGs from stellar dynamics and gravitational lensing
studies (that is, $\fDM\sim 0.3$; \citealt{Capp06}, \citealt{Gerhard},
\citealt{Thomas05}, \citealt{Treu}). In particular, for the NFW families we
found quite high $\fDM$ values (of the order of $\sim 0.66$ for the progenitor)
due to the larger $M_{\mathrm{h}}$ values.

\subsection{From spherical to flat ETGs} \label{sec:flat}
In principle, realistic flat and rotating galaxy models could be constructed with a
Monte-Carlo approach, where all the model parameters are randomly extracted from large ranges, the resulting
models are projected and observed at random orientations, and then checked against the observed
SLs, retaining only those in accordance with observations \citep{LanzCioz}.
This approach is unfeasible here, because the model construction is based on numerical integration (while \citealt{LanzCioz} used
the fully analytical but quite unrealistic Ferrers models), and the computational time of a Monte Carlo exploration
of the parameter space would be prohibitively large. So we solved the problem as follows.

We flatten each spherical progenitor, acting on the axial ratio $q$ and on the scale-lenght
$\reo$ of the stellar density in Eq.~(\ref{eq:rho_*2}), while keeping $\Lr$, $\Upsilon_*$
(and then $M_*$ and $M_{\mathrm{BH}}$), and the DM halo the same.
For given $q$ and $\reo$, the circularized effective radius $\re$ depends on the line-of-sight (l.o.s.) direction, ranging
from $\reo$ (when the model is observed face-on, hereafter FO) to $\sqrt{q}\reo$ (in the edge-on case, EO).
Thus, a request for a realistic model is that $\Lr$ and $\re$ remain consistent with the
observed Size--Luminosity relation, independently of the l.o.s. direction.
In turn, also $\se$ will change due to the flattening, both as a consequence of the choice of
$q$ and $\reo$, and of the l.o.s. inclination.

To include all possible inclination effects, from each spherical progenitor,
we build two sub-families of flat descendants, the FO-built ones and the EO-built ones.
In the first sub-family, $\re$ is the same of the spherical progenitor when the flat model is seen FO;
in the other, $\re$ is the same of the spherical progenitor when the flat model is seen EO.
This implies that $\reo$ may vary: with the decrease of $q$, $\reo$ remains equal to $\re$
of the spherical progenitor in the FO-built case, while $\reo$ increases as $\re/\sqrt{q}$
in the EO-built case. Therefore, in this latter case, there is a consequent expansion
(or size increase) of the galaxy, and a decrease of the galaxy scale density
$\rho_0\propto\sqrt{q}$ in Eq.~(\ref{eq:rho_*2}). On the contrary, in the FO-built sub-family,
there is a density increase as $\rho_0\propto q^{-1}$, as the galaxy is compressed along the $z$-axis.

Then, we compute $\se$ according to the procedure described in Appendix~\ref{sec:code},
for the range spanned by the Satoh parameter $0\leqslant k\leqslant1$.
Since the FO and EO-built models are characterised by different structural
and dynamical properties, we must check that, once observed along arbitrary
inclinations, the models are still consistent with the observed SLs.
For example, a FO-built model, when observed EO, will have an $\re$ smaller than the progenitor,
while an EO-built model will have a larger $\re$ when observed FO.Therefore, only galaxy models that, observed along
the two extreme l.o.s. directions (FO and EO), lie within the observed scatter of $\re$ and $\se$
at fixed $\Lr$ should be retained in our study. Remarkably, all the models constructed with our procedure have been
found acceptable.

The effects of flattening on $\se$ deserve some comments.
For the EO view, $\se$ of the EO-built models decreases for increasing flattening, due to the associated model
expansion; $\se$ further decreases at increasing $k$, as more galaxy flattening
is supported by ordered rotation. 
Also for the FO view, $\se$ of EO-built systems decreases,
but independently of $k$ (affecting
only $\sigma_{\varphi}$, while $\sigma_{\mathrm R}=\sigma_{\mathrm z}$).
In the FO-built models, one would naively expect an increase of $\se$ due to the density
increase (and so to the gravitational potential deepening), but this is not the case: 
even though less severely than for EO-built models, $\se$ still decreases (both
for the FO and EO views). 
The simplest way to explain this behaviour is
to consider the FO flattening of the fully analytical Ferrers ellipsoids \citep{BinTre}.
As the flattening increases, the density raises, the gravitational potential
well deepens, and the vertical force increases, but again the velocity dispersion drops. 
The physical reason, behind the mathematics [e.g., see eqs.~(C4-C11) in \citet{LanzCioz}],
is that stars need less vertical velocity dispersion in order to support the decreased
$z$-axis scale-lenght, and so the FO view $\se$ decreases.
$\se$ decreases less when observed EO, because of the decrease of $\re$, that causes $\se$
to be computed within a smaller area around the galactic centre.
Of course, if the galaxy is not fully velocity dispersion
supported, the decrease of $\se$ for an EO view can be even larger than for the FO
one, since part of the stellar kinetic energy is stored in ordered
motions that do not contribute to $\se$.

\begin{figure}
\includegraphics[width=\linewidth]{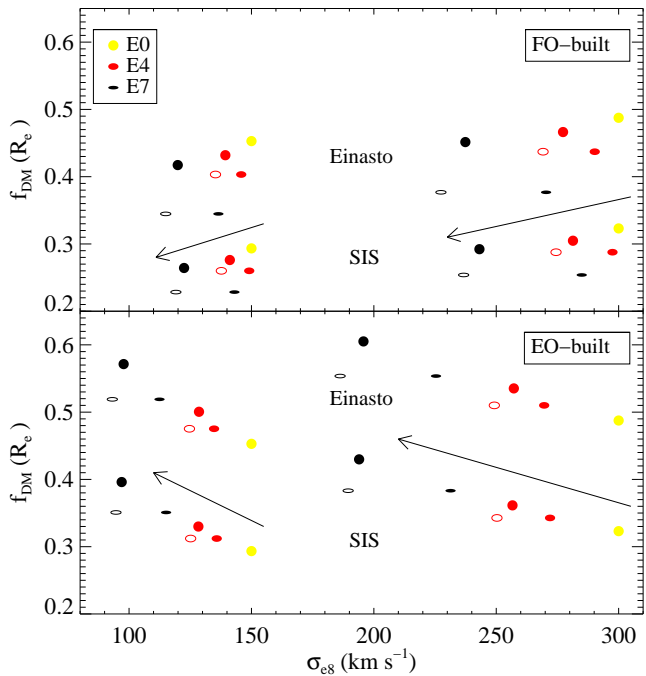}
\caption{Dark matter fraction $\fDM$ as a function of the shape parameter $q$ for the SIS (lower symbols) and the Einasto (upper symbols) DM halo models,
for two families with $\se=150$ and 300 $\kms$ for the spherical progenitors: the FO-built sub-families are in the top panel,
while the EO-built ones are in bottom panel. The yellow, red, and black colours refer
to the E0, E4, E7 model galaxies, respectively; symbols are filled for $k=0$, and empty for $k=1$.
The shape of the symbols (round or elliptical) indicates the FO or the EO 
view of a given model. See Sect.~\ref{sec:obs} and \ref{sec:flat} for more details.
The arrows indicate the trends of change of $\fDM$ for increasing flattening, and separate the symbols of the two DM profiles.}
\label{fig:fDM}
\end{figure}
These general results, obtained for realistic models, about the variation of $\se$
in flat and rotating galaxies of fixed stellar mass, show that
some caution should be exercised
when using simple dynamical mass estimators based on the velocity dispersion measured in the
central regions of galaxies.
This point is particularly relevant for studies of the hot haloes properties, that notoriously mainly
depend on the galaxy mass (CDPR, S13).

As a further test of the models, we also
calculated the parameter $\lambda_R$, introduced by \citet{SauronIX} and related to the mean amount of stellar rotational support.
Our $\lambda_R$ radial profiles, even for the $k=1$ case, are in good agreement with the profiles
in the ATLAS$^{\mathrm{3D}}$ sample of ETGs (fig.~5 in \citealt{Atlas3DIII}), for each galaxy ellipticity.
Finally, our method of definition of the DM halo implies a constant $M_{\mathrm h}/M_*$ ratio within each
family, but not a constant $\fDM$, that depends on $q$ (through the
variation that $q$ imposes to $\rho_*$), as one can see in Fig.~\ref{fig:fDM}
for the SIS and the Einasto families. Note that $\fDM$ can decrease or increase
with $q$ depending on the construction mode: the increase of the stellar density in the FO-built
models results in lower $\fDM$, since the DM halo is fixed; the reverse is true for the EO-built models.
Moreover, the Einasto models have always higher DM fractions than the corresponding
SIS ones, due to the steepness of the SIS profile at small radii, requiring
less DM to reproduce the chosen value of $\se$.

\section[]{Results} \label{sec:res}
Having built a large set of realistic galaxy models, consistent with the observed SLs and
with DM haloes in agreement with current expectations, we can now study
the effects of flattening and rotational support on the temperatures of the models, defined in Sect.~\ref{sec:temp}
and summarized in Table~\ref{tab:allparam}, together with the main parameters characterizing the models.
We next compare these temperatures with the observed X-ray properties of a sample of ETGs, extending to
flat and rotating models the analysis carried out by P11.
In this Section, we take into account also the effect of $\alpha$, that parametrizes the degree of thermalisation
of the ordered motions. 

\begin{table*}
\caption{Summary of all parameters and temperatures.}
\begin{tabular}{cl}
\toprule
Symbol   & Meaning   \\
\midrule                                    
$q$      & Intrinsic axial ratio of the stellar distribution: $0.3\leqslant q\leqslant1$    \\
$k$      & Satoh parameter, controls the amount of galaxy rotation: $0\leqslant k\leqslant1$   \\
$\gth$   & Degree of thermalisation of the ordered stellar motions $\Bld{v}$ [Eq.~(\ref{eq:gth})]  \\
$\alpha$ & Scaling factor between the ISM velocity and the stellar streaming motions ($\Bld{u}=\alpha\Bld{v}$), with $0\leqslant \alpha\leqslant1$ ;
$\gth=(\alpha-1)^2$ \\
$\Tinj$  & Temperature equivalent of the thermalisation of the stellar motions and of the kinetic energy of SNe Ia events, for the unit mass of\\
         & injected gas [Eq.~(\ref{eq:tinj})] \\          
$\Tsn$   & Contribution to $\Tinj$ due to SNIa events [Eq.~(\ref{eq:tsn})]; it is regulated by a factor $\eta<1$ ($\eta=0.85$ is generally adopted)  \\         
$T_*$    & Contribution to $\Tinj$ due to stellar motions, defined by Eq.~(\ref{eq:tstar}): $T_*=T_{\sigma}+\gth T_{\mathrm{rot}}$\\           
$T_{\sigma}$& Contribution to $T_*$ due to stellar random motions, defined by Eq.~(\ref{eq:tsig})     \\
$T_{\mathrm{rot}}$& Contribution to $T_*$ due to stellar ordered motions defined by Eq.~(\ref{eq:trot})    \\           
$\Tgp$   & Temperature equivalent of the change in gravitational energy of the injected gas, when flowing to the galactic centre [Eq.~(\ref{eq:egr+})]   \\
$\Tgm$   & Temperature equivalent of the energy required to extract the unit mass of injected gas from the galaxy [Eq.~(\ref{eq:tgr-})]. If the injected gas\\
         & rotates, then $\Tgm(\alpha)=\Tgm-\alpha^2 T_{\mathrm{rot}}$\\
$\Tsub$  & Mass averaged, subsonic escape temperature for the injected gas [Eq.~(\ref{eq:tsub})]. It represents a fiducial upper limit to the observed\\
         & temperature of outflows; $\Tsub=3/5\,\Tgm$\\
\bottomrule
\end{tabular}
\label{tab:allparam}
\end{table*}

\subsection[]{The effects of shape and stellar streaming motions on the model temperatures}\label{sec:eff}
\begin{figure}
\includegraphics[width=\linewidth]{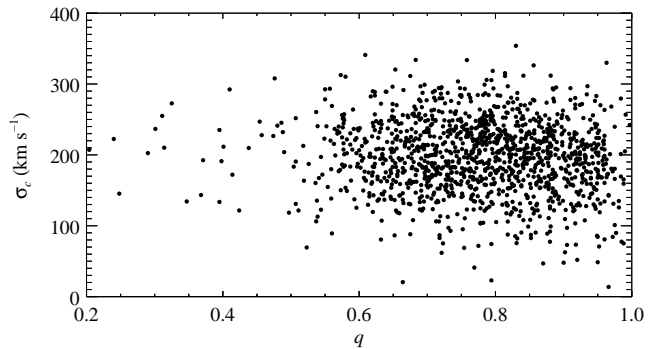}
\caption{Central stellar velocity dispersion as a function of the shape parameter $q$
for a sample of $\approx$ 1400 nearby ($z<0.05$) ETGs drawn from the SDSS DR4
(data taken form \citealt{Nair}).}
\label{fig:nair}
\end{figure}

We explore here how $T_*$, $\Tinj$ and $\Tgm$ depend on
$(q, k,\alpha)$, i.e., galaxy flattening, rotational support, and degree of thermalisation of
ordered rotation. Three values of $q= (1,0.6,0.3)$ are considered, that
cover ETG morphologies from the spherical (E0) to the flattest ones
(E7). The choice of the intermediate value $q=0.6$ (corresponding to an E4) is motivated by the majority of ETGs
having $0.55\lesssim q\leqslant 1$ (see Fig.~\ref{fig:nair}). 
Thus, the E7 models correspond to rare objects and represent quite an extreme behaviour,
whereas the most common ETGs correspond to models with $q$ between $1$
and $0.6$. We also consider the two extreme values of $k$: fully velocity
dispersion supported systems ($k=0$), and isotropic rotators ($k=1$);
and three values of $\alpha= (0,0.5,1)$, in which respectively the pre-existing ISM
has a null rotational velocity (all the stellar kinetic energy,
including that of rotational motions, is thermalised,
$\gth=1$), or rotates with half the velocity of the stars (then
$\gth=0.25$), or has the same velocity as the stars (then no
ordered stellar kinetic energy is thermalised, $\gth=0$, see Sect.~\ref{sec:temp}).

\begin{figure*}
\includegraphics[width=0.48\linewidth]{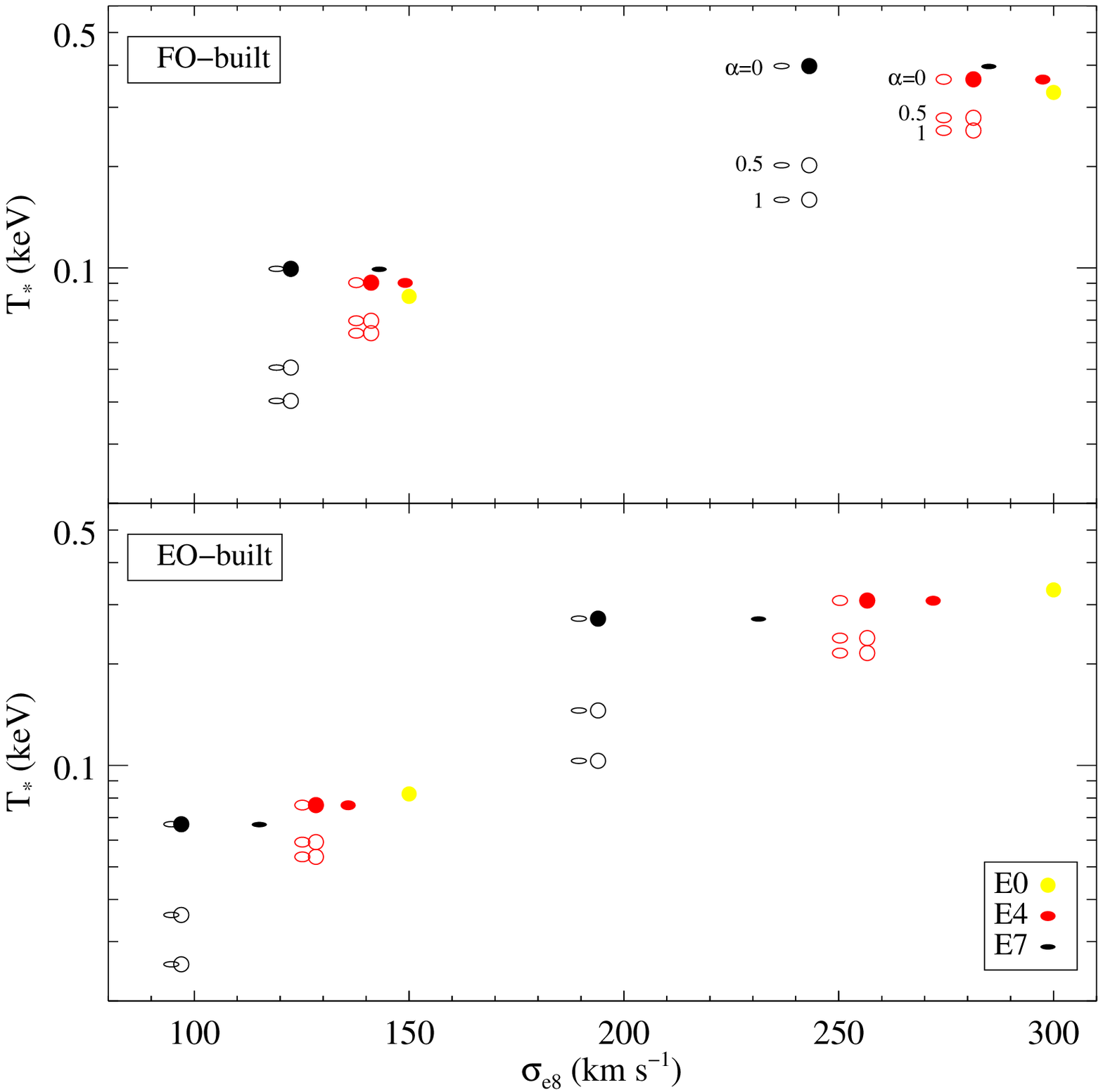}
\hskip 0.5truecm
\includegraphics[width=0.48\linewidth]{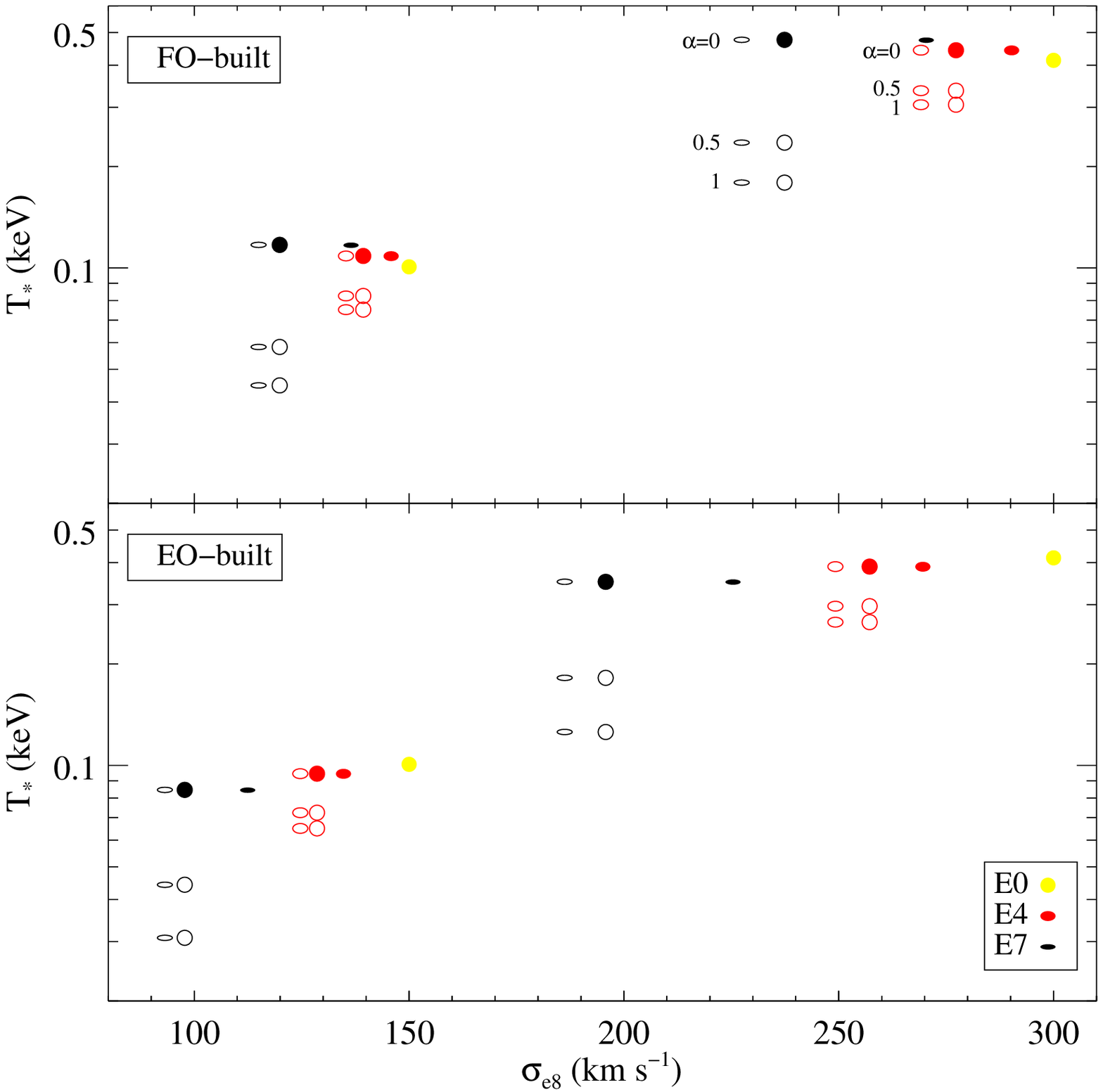}
\caption{Distribution of $T_*$ for models derived from two spherical progenitors (yellow circles) of $\se=150$ or 300 $\kms$,
with SIS (left panel) or Einasto DM halo (right panel). Colours refer to the \textit{intrinsic} flattening of the descendants:
E4 (red) and E7 (black). At given intrinsic flattening (i.e., at fixed colour), the shape of the symbols indicates
the l.o.s. inclination: FO (circle) or EO view (ellipse). Filled symbols refer to $k=0$, empty symbols to $k=1$.
Three values of $\alpha$ are considered ($\alpha=0, 0.5, 1$), as indicated for the FO-built sub-families of the $\se$=300 $\kms$ progenitor.
When $k=1$, the increase of $\alpha$ decreases $T_*$ at any $\se$, from the value coincident with 
the non-rotating case ($k=0$) when $\alpha=0$, down to values that are lower for larger flattenings, when $\alpha=0.5$ and 1.
In each sub-family, models with same intrinsic shape have the same $T_*$, independent of the FO or EO view, for fixed $(k,\alpha)$.}
\label{fig:TstSIS}
\end{figure*}

\begin{figure*}
\includegraphics[width=0.48\linewidth]{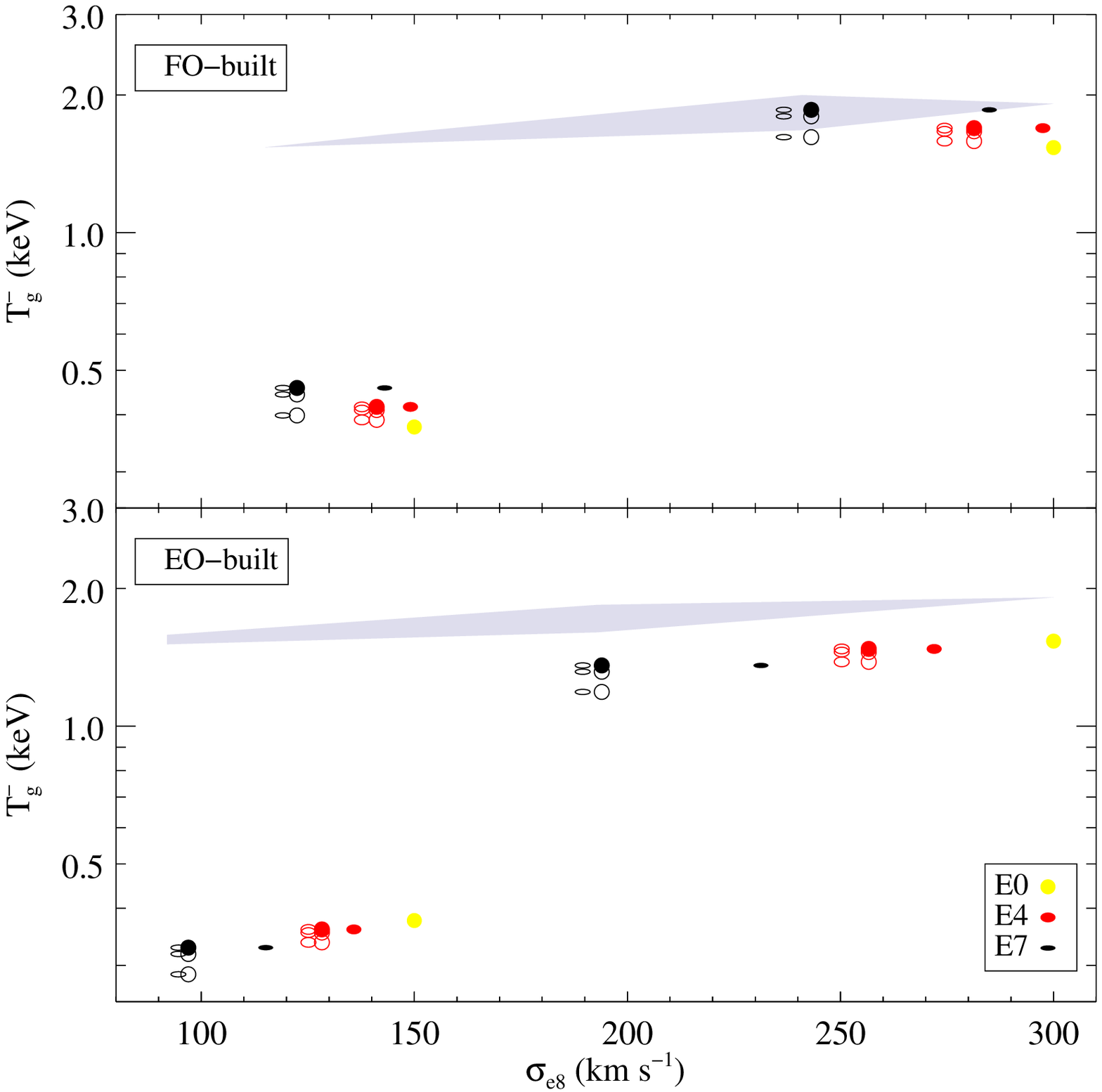}
\hskip 0.5truecm
\includegraphics[width=0.48\linewidth]{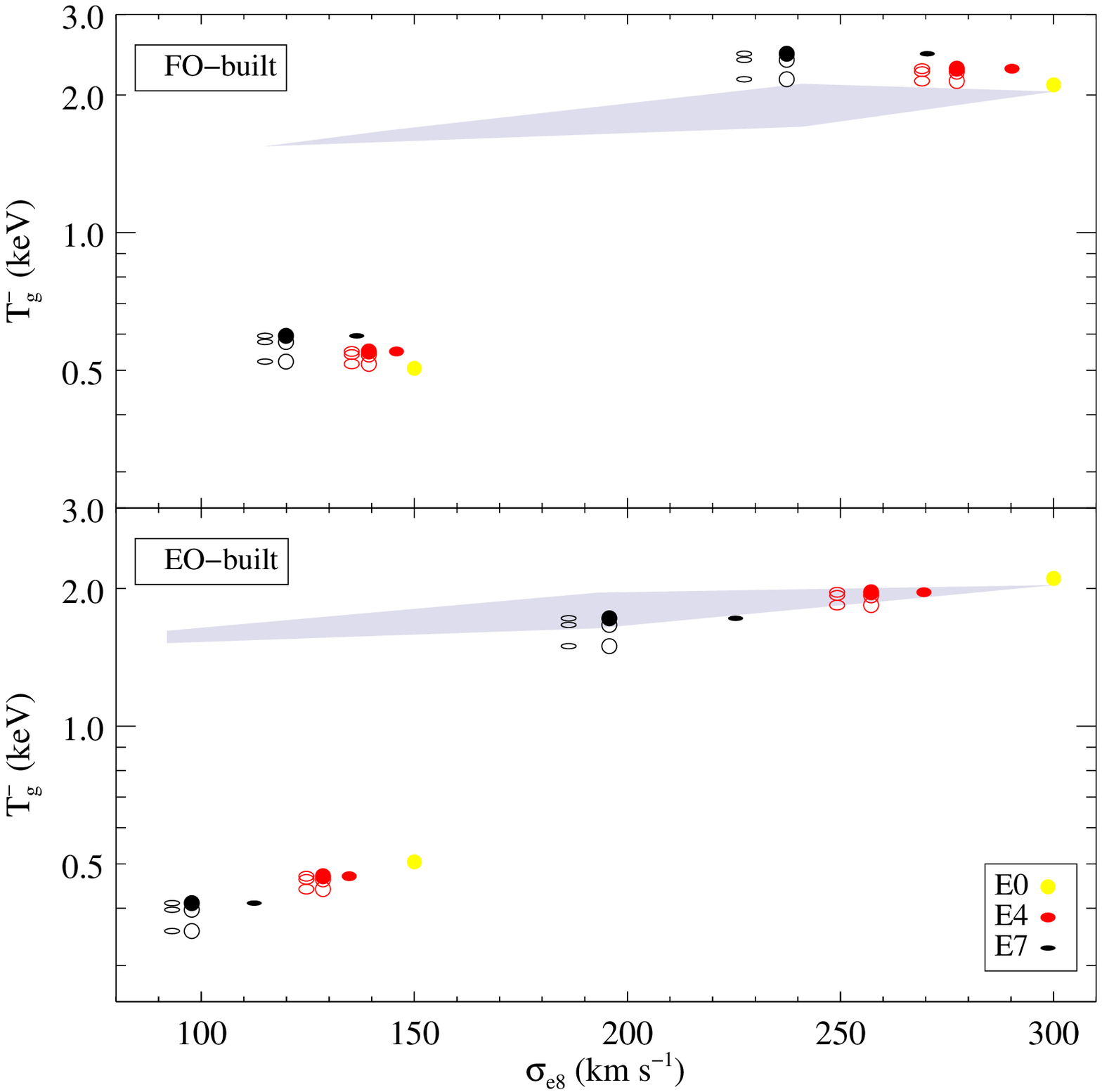}
\caption{$\Tgm$ for the same models in Fig.~\ref{fig:TstSIS} (SIS and Einasto DM halo in the left and right panels, respectively);
the same notation applies. The shaded area shows the range spanned by $\Tinj$, considering
the variations in $T_*$ in Fig.~\ref{fig:TstSIS}, for $\eta=0.85$.}
\label{fig:TinjSIS}
\end{figure*}

Figure~\ref{fig:TstSIS} shows $T_*$ for various descendants of spherical progenitors with 
$\se=150$ and 300 $\kms$ (yellow circles), for two different DM haloes (SIS and Einasto in the left
and right panels, respectively). As anticipated in Sect.~\ref{sec:flat},
a major effect of flattening is the decrease of $\se$ of the descendants (with respect to
the progenitor), that are then displaced on the left of their respective progenitor, in a way
proportional to the flattening level
\footnote{In some works, instead of $\se$, the observations are used to measure
the quantity $\Vrms=\sqrt{\sigma_{\mathrm P}^2+V_{\mathrm P}^2}$, averaged within a central
aperture (i.e., $\re/8$) by weighting with the surface brightness (see Appendix~\ref{sec:code}).
For a chosen shape of the stellar distribution, $\Vrms=\sigma_{\mathrm P}$ if $k=0$,
and whenever the galaxy is seen face-on. For any view, it can be shown that, for axisymmetric
stellar distributions where the Satoh $k$-decomposition is adopted, $\Vrms$ is independent of $k$;
thus $\Vrmse$ has a different behavior than $\se$, that is slightly lower for $k=1$ than for $k=0$, for the edge-on view
(Figs.~\ref{fig:TstSIS} and \ref{fig:TinjSIS}).}
(arrows in Fig.~\ref{fig:fDM}, see also Figs.~\ref{fig:TstSIS} and \ref{fig:TinjSIS}).

The trend of $T_*$ with a pure change of shape in fully velocity dispersion supported models
($k=0$), is due to the specific flattening procedure (see Sect.~\ref{sec:flat}).
In the FO-built sub-families (Fig.~\ref{fig:TstSIS}, top panels), flatter models are more concentrated than rounder ones, while
in the EO-built sub-families (bottom panels), they are more extended and diluted.
Thus pure flattening produces a different effect on $T_*$: in the FO-built cases $T_*$ increases (as $\Tgm$; see below),
whereas in the EO-built cases $T_*$ decreases.
Therefore, we conclude that real flat galaxies can be either more or less bound
than spherical galaxies of the same mass, depending on their mass concentration.
Overall, however, the variation in $T_*$ for both sub-families is not large: the maximum variation, 
from the progenitor to the E7 model, is an increase of $\sim 19$ percent 
for the FO-built cases, and a decrease of $\sim 18$ percent for the EO-built ones.

A larger effect on $T_*$ can instead be due to the presence of significant
rotational support (empty symbols in Fig.~\ref{fig:TstSIS}), if not thermalised. In fact, when
$k=1$, but $\alpha=0$ ($\gth=1$), the whole stellar kinetic energy, including the streaming one, is thermalised, and the $T_*$
values are coincident with those of the non-rotating case (full
symbols), for the same galaxy shape. In the other cases of
$\alpha\neq 0$, the rotational support always acts in the sense of
reducing $T_*$, and the flatter the shape, the larger can be the reduction. The
strongest reduction of $T_*$ is obtained for an isotropic rotator
($k=1$) E7 model, if the gas ejected from stars retains the same
stellar streaming motion ($\alpha=1$, $\gth=0$): for the
FO-built case, $T_*$ drops by $\sim50$ percent with respect to the E0 model, and
by $\sim60$ percent with respect to the same E7 model with the ordered streaming
motions fully thermalised ($\alpha=0$). For the EO-built case, $T_*$
drops by $\sim70$ percent with respect to the E0 model, and by $\sim60$ percent with respect
to the same E7 model with $\alpha=0$. These percentages are obviously
extreme values; the $T_*$ reduction is lower for milder flattenings, and for $k$ and
$\alpha$ values smaller than 1. For a fixed galaxy shape,
all possible $(k,\alpha)$ combinations fill a sort of
triangular area on the $(\se,T_*)$ plane, identifiable by linking the
symbols of a given $q$ (colour). Clearly, the rounder the galaxies, the
weaker the effect of $k$ and, consequently, of $\alpha$ variations.
All the above effects are independent of the galaxy luminosity (mass),
and both the $\se=300$ and 150 $\kms$ families
show the same (rescaled) behaviour in the $(\se,T_*)$ plane.

The trends described above are independent of the specific DM halo profile:
models with an Einasto DM halo (right panel) show the same pattern as those with a SIS DM halo (left panel), just with a
different normalisation due to the larger total DM content (see Table~\ref{tab:params}).
Similar results hold also for the Hernquist and NFW DM haloes.

In Fig.~\ref{fig:TinjSIS} we plot $\Tgm$ for the same families in
Fig.~\ref{fig:TstSIS}. Similarly to what happens for $T_*$, for
fully velocity dispersion supported models, $\Tgm$ gets larger with
flattening for the FO-built sub-family ($\sim 21$ percent), while it
decreases for the more diluted EO-built models ($\sim 12$ percent).
Stellar streaming, when $\alpha >0$,
acts in the sense of making the gas less bound, due to the centrifugal
support of the injected gas, and then $\Tgm$ decreases with increasing $\alpha$, at any
fixed flat shape. This effect is maximum when $\alpha =1$ and the gas
rotates as the stars. However, the decrease in $\Tgm$ due to galaxy
rotation is lower than obtained for $T_*$: for both sub-families, $\Tgm$ 
drops at most by $\sim13$ percent (for the E7 models), between the two extreme
cases of $\alpha=0$ and $\alpha=1$.
This produces that, in the FO-built case, $\Tgm$ keeps always larger than for
the progenitor when $q$ decreases, even for $k=\alpha =1$, while $T_*$ of rotating
galaxies could become significantly lower than for the
progenitor. Note that there are two compensating effects from stellar streaming when
$\alpha\neq 0$: the stellar heating is lower than for $k=0$, but the
gas is also less bound. 
Also these results are independent of the
DM halo profile, as can be judged from the right panel of Fig.~\ref{fig:TinjSIS}, that
refers to the same models of the right panel of Fig.~\ref{fig:TstSIS}.

Since the flatter is the galaxy, the more it can be rotationally supported (and the more is rotationally
supported, the larger is the effect of a corotating ISM), the effect of rotation is dependent
on the degree of flattening, and thus it may prove difficult to disentangle observationally the
two distinct effects due to shape and kinematics.
On the theory side, we recall that a simplifying assumption made here is 
that $\Bld{u}=\alpha\Bld{v}$, and $\gth=(\alpha -1)^2$,
while in reality the kinematical difference between stars and pre-existing gas may
be more complex, as the extent of thermalisation; only numerical simulations 
will be able to establish what are the net effects on the gas evolution of stellar streaming motions (see
\citealt{ProcAnd,Andrea}).

Figure~\ref{fig:TinjSIS} finally shows the well known fact that the
contribution from SNIa's dominates the gas injection energy, since
$\Tinj >> T_*$. This contribution (i.e., $\Tsn$) is independent of
galaxy mass, which results into lower-mass models having $\Tgm$ far
lower than $\Tinj$, and $\Tgm$ reaching $\Tinj$ for 200 $\kms\lesssim\se\lesssim250$
$\kms$, depending on the DM profile. Galaxies with $\se\lesssim 200$ $\kms$
consequently are more prone to an outflow, and then to have
a low hot gas content, as already suggested in the past by numerical
simulations and by observations (CDPR, \citealt{Sarazin01,David,Pel07,Trinchieri}).
These findings are based on the assumption of a high thermalisation
efficiency for SNIa's ($\eta=0.85$), and the quoted $\se$ critical values become lower for lower
$\eta $ values (as indicated by, e.g., \citealt{Thornton}), that decrease $\Tinj$.
Variations in the dark matter may alter the $\Tgm$ values, but small
changes in $\Tgm$ are found here for differences in the DM profile,
and possible variations in the total DM amount cannot be very large,
given the constraints from dynamical modelings within $\re$, and from
cosmological simulations (taken into account here, Sect.~\ref{sec:obs}).
Indeed, the results for the Hernquist families are essentially
identical to what we have shown for the Einasto halo, whereas for the NFW profile the trends are the same,
but all the temperatures are shifted to higher values, due to their larger amounts of DM.

\begin{table*}
\caption{Observed Properties of the ETG Sample with X-ray properties for the hot gas from Chandra observations.}
\begin{tabular}{ccccccccccc}
\toprule
Name     &$d$    &$\log(\Lk)$&$\kb\Tx$&$\Lx$           &$\Vmax$ &$\se$   &$\Vmax/\se$&Ref.                 &Type           &$q$    \\
         &(Mpc)  &$(\Lks)$   &(keV)   &$(10^{40}\ergs)$&$(\kms)$&$(\kms)$&           &                     &RC3            &2MASS  \\
(1)      & (2)   & (3)       & (4)    & (5)            & (6)    & (7)    & (8)       &(9)                  &(10)           & (11)  \\
\midrule                                                                                                                                     
NGC 720  & 27.6  &11.31  & 0.54   &5.06            &100   &  241  &   0.41    &Binney et al. 1990     & E5            &0.55   \\
NGC 821  & 24.1  &10.94  & 0.15   &2.13$\e{-3}$    &120   &  200  &   0.60    &Coccato et al. 2009    & E6?           &0.62   \\
NGC1023  & 11.4  &10.95  & 0.32   &6.25$\e{-2}$    &250   &  204  &   1.23    &Noordermeer et al. 2008& SB0           &0.38   \\
NGC1052  & 19.4  &10.93  & 0.34   &4.37$\e{-1}$    &120   &  215  &   0.56    &Milone et al. 2008     & E4            &0.70   \\
NGC1316  & 21.4  &11.76  & 0.60   &5.35            &150   &  230  &   0.65    &Bedregal et al. 2006   & SAB0          &0.72   \\
NGC1427  & 23.5  &10.82  & 0.38   &5.94$\e{-2}$    & 45   &  171  &   0.26    &D'Onofrio et al. 1995  & cD            & --      \\
NGC1549  & 19.6  &11.20  & 0.35   &3.08$\e{-1}$    & 40   &  210  &   0.19    &Longo et al. 1994      & E0-1          &0.90   \\
NGC2434  & 21.5  &10.84  & 0.52   &7.56$\e{-1}$    & 20   &  205  &   0.10    &Carollo et al. 1994    & E0-1          &0.98   \\
NGC2768  & 22.3  &11.23  & 0.34   &1.26            &195   &  205  &   0.95    &Proctor et al. 2009    & E6            &0.46   \\
NGC3115  &  9.6  &10.94  & 0.44   &2.51$\e{-2}$    &260   &  239  &   1.09    &Fisher 1997            & S0            &0.39   \\
NGC3377  & 11.2  &10.45  & 0.22   &1.17$\e{-2}$    & 97   &  144  &   0.67    &Simien et al. 2002     & E5-6          &0.58   \\
NGC3379  & 10.5  &10.87  & 0.25   &4.69$\e{-2}$    & 60   &  216  &   0.28    &Weijmans et al. 2009   & E1            &0.85   \\
NGC3384  & 11.5  &10.75  & 0.25   &3.50$\e{-2}$    &150   &  161  &   0.93    &Fisher 1997            & SB0           &0.51   \\
NGC3585  & 20.0  &11.25  & 0.36   &1.47$\e{-1}$    &200   &  198  &   1.01    &Fisher 1997            & E6            &0.63   \\
NGC3923  & 22.9  &11.45  & 0.45   &4.41            & 31   &  250  &   0.12    &Norris et al. 2008     & E4-5          &0.64   \\
NGC4125  & 23.8  &11.35  & 0.41   &3.18            &150   &  227  &   0.66    &Pu et al. 2010         & E6 pec        &0.63   \\
NGC4261  & 31.6  &11.43  & 0.66   &7.02            & 50   &  300  &   0.17    &Bender et al. 1994     & E2-3          &0.86   \\
NGC4278  & 16.0  &10.87  & 0.32   &2.63$\e{-1}$    & 60   &  252  &   0.24    &Bender et al. 1994     & E1-2          &0.93   \\
NGC4365  & 20.4  &11.30  & 0.44   &5.12$\e{-1}$    & 80   &  245  &   0.33    &Surma et al. 1995      & E3            &0.74   \\
NGC4374  & 18.3  &11.37  & 0.63   &5.95            & 60   &  292  &   0.21    &Coccato et al. 2009    & E1            &0.92   \\
NGC4382  & 18.4  &11.41  & 0.40   &1.19            & 70   &  187  &   0.37    &Fisher 1997            & SA0           &0.67   \\
NGC4472  & 16.2  &11.60  & 0.80   &18.9            & 73   &  294  &   0.25    &Fisher et al. 1995     & E2            &0.81   \\
NGC4473  & 15.7  &10.86  & 0.35   &1.85$\e{-1}$    & 70   &  192  &   0.36    &Emsellem et al. 2004   & E5            &0.54   \\
NGC4526  & 16.9  &11.20  & 0.33   &3.28$\e{-1}$    &246   &  232  &   1.06    &Pellegrini et al. 1997 & SAB0          &0.43   \\
NGC4552  & 15.3  &11.01  & 0.52   &2.31            & 17   &  268  &   0.06    &Krajnovi\'{c} et al. 2008& E0-1        &0.94   \\
NGC4621  & 18.2  &11.16  & 0.27   &6.08$\e{-1}$    &140   &  225  &   0.62    &Bender et al. 1994     & E5            &0.65   \\
NGC4649  & 16.8  &11.49  & 0.77   &11.7            &120   &  315  &   0.38    &Pinkney et al. 2003    & E2            &0.81   \\
NGC4697  & 11.7  &10.92  & 0.33   &1.91$\e{-1}$    &115   &  174  &   0.66    &De Lorenzi et al. 2008 & E6            &0.63   \\
NGC5866  & 15.3  &10.95  & 0.35   &2.42$\e{-1}$    &210   &  159  &   1.32    &Neistein et al. 1999   & SA0           &0.42   \\
\bottomrule
\end{tabular}
\flushleft
Notes: (1) Galaxy name. (2) Distances from BKF \citep{Tonry}. (3) Logarithm of the $K$ band luminosity, assuming $K_{\odot}=3.33$ mag, taken from BKF.
$(4)-(5)$ Hot gas temperature and the $0.3-8$ keV gas luminosity, from BKF.
$(6)-(8)$: maximum velocity of rotation, stellar velocity dispersion, as the luminosity-weighted average within an aperture of radius $\re/8$ (from P11),
and their ratio. (9) References for $\Vmax$ in column (6). (10) Morphological type from RC3. (11) Axial ratio in the $K_s$ band, from 2MASS.
\label{tab:ETGs}
\end{table*}

Finally we comment on the preliminary investigation about the role of flattening and rotation
on the global energetics of the ISM in CP96. They built fully analytical axisymmetric two-component galaxy models,
where the stellar and dark mass distributions were described by the Miyamoto--Nagai
potential-density pair. They varied the shape of both the stellar component and
the dark matter halo from flat to spherical, and the amount of azimuthal ordered
motions through the Satoh $k$-decomposition.
Their conclusion was that,
for quite round systems, flattening can have a substantial effect in reducing
the binding energy of the hot gas, contrary to galaxy rotation that seemed to have a negligible role.
The opposite was suggested for very flat systems.
These results were confirmed by 2D hydrodynamical simulations \citep{DercCio98}.
We stress here that the models in CP96, while capturing the main effects of flattening and
rotation on the global energetics of the hot haloes of ETGs, were not tailored to reproduce in detail
the observed properties of real ETGs.
Our current findings, based on more realistic galaxy models, reveal a more complicated situation,
where the flattening importance on the ISM status is mediated by the amount of rotation
and its specific thermalisation history. Remarkably enough, however, when flattening a
two-component Miyamoto--Nagai model following the same procedure here adopted
(constant $M_*$, $\re$ and $M_{\mathrm h}$), its $\Tgm$ remains close to that of its spherical
progenitor (i.e., the model moves almost parallel to the solid lines in fig.~2 in CP96).

\subsection{Comparison with observed ETGs properties in the X-rays}\label{sec:compwithdata}
\begin{figure*}
\includegraphics[width=0.48\linewidth]{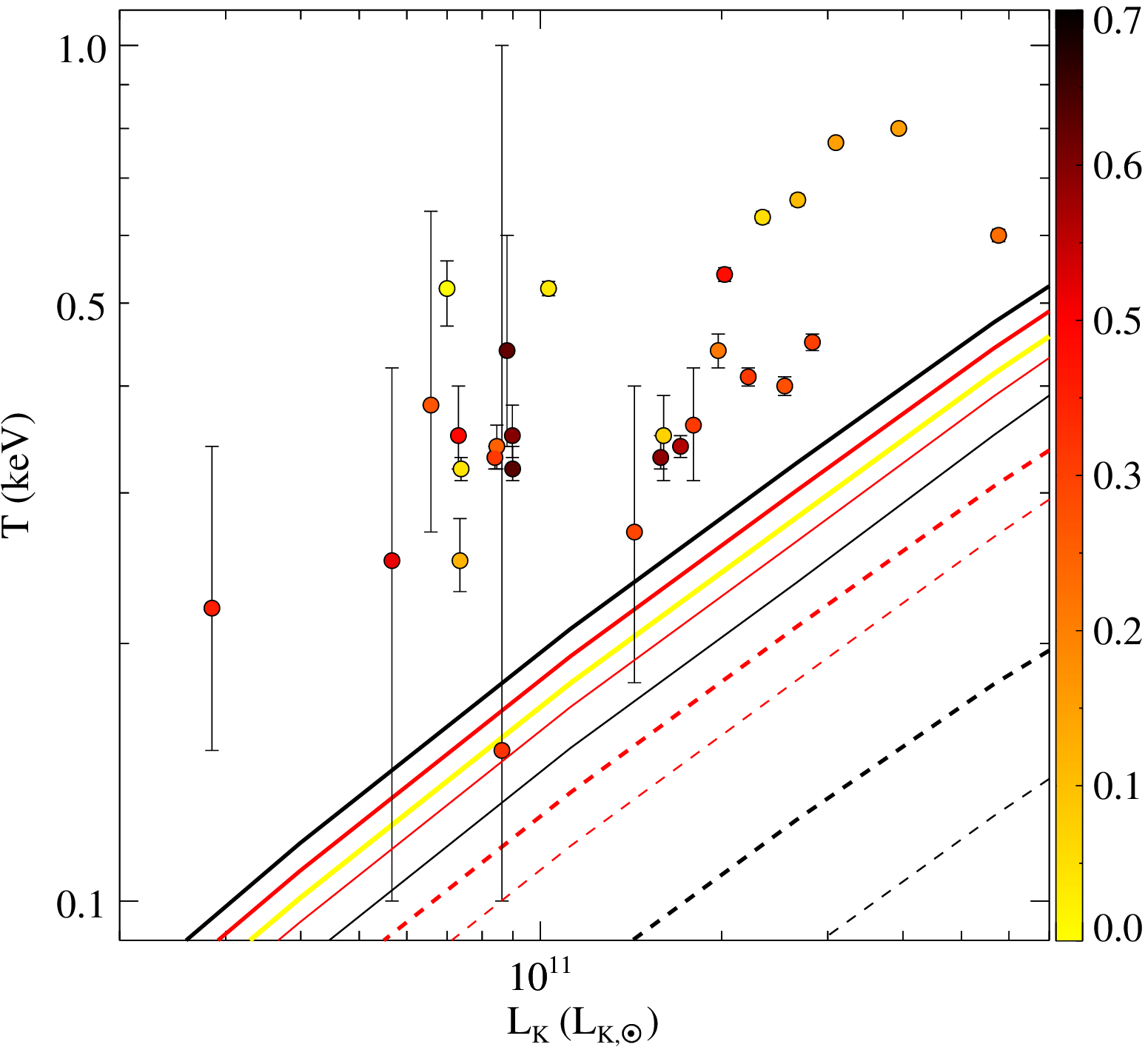}
\hskip 0.5truecm
\includegraphics[width=0.48\linewidth]{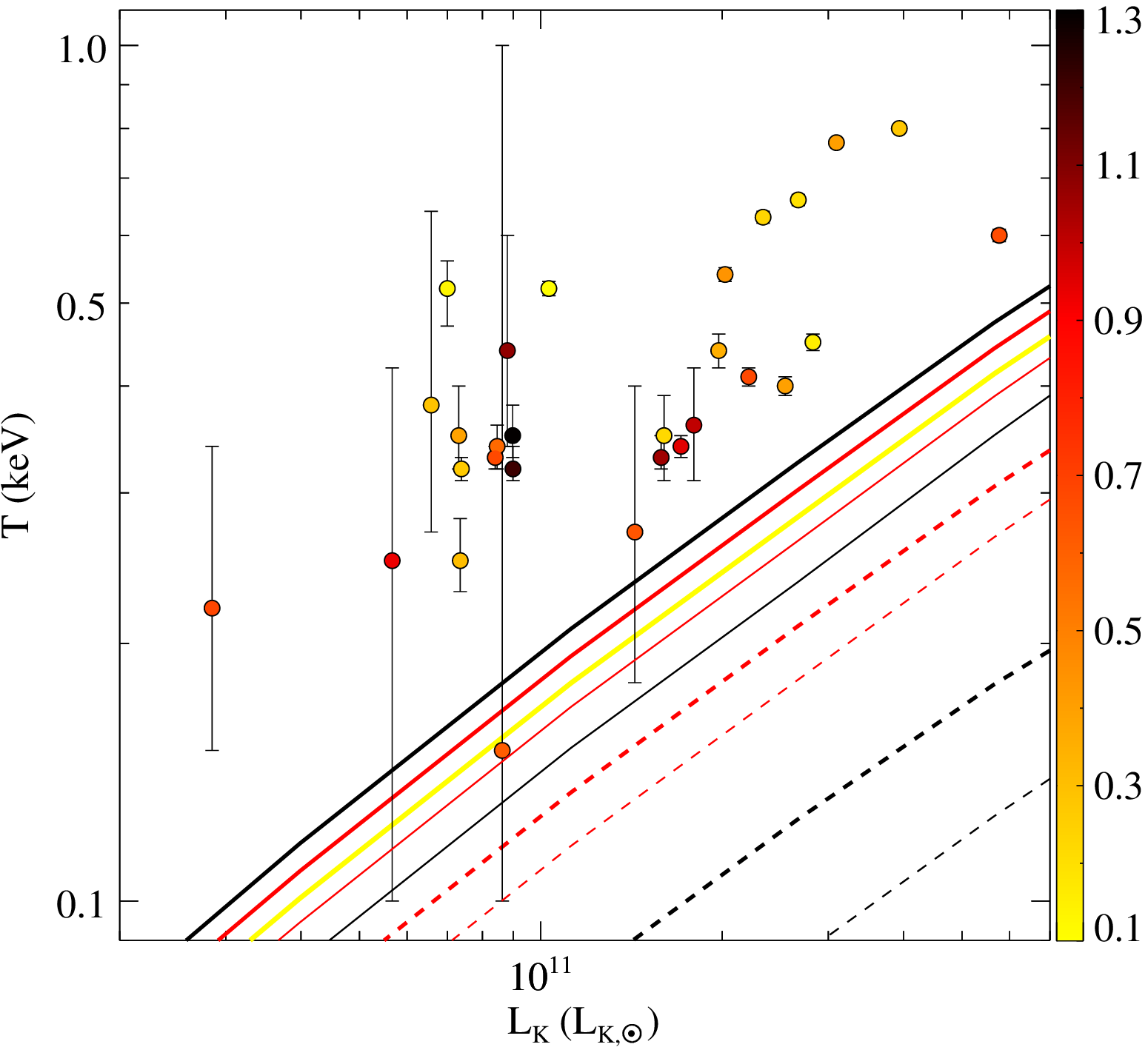}
\caption{$T_*$ (lines) for all the progenitors in Tab.~\ref{tab:params} with the Einasto halo,
and their descendants (with $\alpha=0$ or $\alpha=1$),
and the observed $\Tx$ (circles), for the ETGs in the BKF sample, as a function of $\Lk$. 
The yellow line refers to the progenitor ETGs, the red lines to the E4 shape, the black ones to the E7 shape;
lines are solid for $k=0$, and dashed for $k=1$ and $\alpha=1$. For each colour, lines
representing the case of $k=1$ and $\alpha=0$ are coincident with the solid lines.
Thick and thin lines refer to the FO and EO-built models, respectively.
Left panel: the colour-coding for the observed ETGs indicates their ellipticity $\epsilon=1-q$, 
as measured in the Ks-band, from 2MASS (see Tab.~\ref{tab:ETGs}),
and is calibrated as for the previous figures [i.e., to be yellow for the E0 ($q=1,\epsilon=0$), red for the E4 ($q=0.6,\epsilon=0.4$), 
and black for the E7 ($q=0.3,\epsilon=0.7$); see the colour bar on the right for the $\epsilon$ of the other colours].
Right panel: the colour-coding indicates the rotational support $\Vmax/\se$. Note that, for an EO view, the models of
the dashed lines would have $\Vmax/\se=0.90-0.95$ for the E4 case, and $\Vmax/\se=1.5-1.6$ for the E7 case.}
\label{fig:tstobs}
\end{figure*}

Now we compare our estimates for $T_*$ and $\Tgm$ with observed
temperatures $\Tx$ and gas content, respectively, for the BKF sample.
Observed X-ray and $K$-band luminosities, central stellar velocity dispersions, rotation
velocities, and galaxy shape are given in Tab.~\ref{tab:ETGs}.
Figure~\ref{fig:tstobs} shows observed (points) and
model temperatures (lines) versus $\Lk$, with $\Tx$ plotted with a
different colour reflecting the ETG shape and rotational support. In
order to make the comparison between models and observations more
consistent, the $\Lk$ of the models have been calculated using the
mean $\Lk/\Lv=3.4$ of the ETGs in this sample. When comparing observed and model temperature
values at fixed $\Lk$, the mass is the same for all the models, and it
should be roughly so also for the observed ETGs. At fixed $\Lk$, then,
the $T_*$ variation due to shape and stellar streaming is obtained from all the $T_*$ values
of the descendants of a progenitor (we consider the Einasto models of Fig.~\ref{fig:TstSIS}), regardless of the FO or EO view.
When comparing observed and model temperatures as a function of $\se$ instead,
at any $\se$ the mass could be different. Figure~\ref{fig:tstobs} shows again
how the effect of rotation can be potentially stronger than that of shape
(as already indicated by Fig.~\ref{fig:TstSIS}): moderately
larger or smaller $T_*$ can be produced by flattening, depending on
the way it is realized, while $T_*$ can be much lowered by galactic
rotation. Thus, at fixed $\Lk$, the largest variation in $T_*$ with
respect to the spherical case (yellow line) does not
come from a variation of shape, but is a decrease of $T_*$ due to
rotation with $\alpha=1$ (no thermalisation).

In Fig.~\ref{fig:tstobs} all $\Tx$ are larger than $T_*$, yet much closer
to $T_*$ than to $\Tinj\sim1.5-2$ keV (see Fig.~\ref{fig:TinjSIS}). 
This may be evidence of two facts: either the SNIa's thermalisation $\eta$ 
is low, or the gas flows establish themselves at a temperature close to the virial temperature\footnote{The
integrals in the definitions of $T_{\sigma}$ and $T_{\mathrm{rot}}$ (Eqs. 3 and 4) are also used to compute 
the total kinetic energy of the stellar motions that enters the virial theorem for the stellar component; thus the 
mass weighted temperature $T_*$, with $\gth=1$, is often referred to as the gas virial temperature.} of the 
galaxy, and most of the SNIa's input is spent in cooling (in gas-rich ETGs), or in lifting the gas from the 
potential well, and imparting bulk velocity to the outflowing gas (in
gas-poorer ETGs; see P11 for a quantification of these effects, and \citealt{Tang09} and \citealt{LiWang}
for addressing also other solutions to this problem). A combination of the two explanations may also be at place, of course.
In any case, the proximity of the observed $\Tx$ values to $T_*$ provides an empirical evidence of the 
importance of the study of $T_*$ and its variations; it would not have been so,
if we had found $\Tx$ to be closer to $\Tinj$.

Going now into the question of whether possible effects from shape and
stellar streaming are apparent on $\Tx$, Fig.~\ref{fig:tstobs}
shows a mild indication that flatter shapes and
more rotationally supported ETGs tend to show a
lower $\Tx$, with respect to rounder, less rotating ETGs.
This is similar to the recent result by S13, that fast rotators seem to be confined to lower
temperatures than slow rotators. S13 suggested that in ETGs with a larger degree of rotational
support, the kinetic energy associated with the stellar ordered motions
may be thermalised less efficiently. Preliminary results of hydrodynamical simulations
seem to indicate that this is the case \citep{ProcAnd,Andrea}.
Based on the analysis of Sect.~\ref{sec:eff}, we can propose an explanation for the trends
in Fig.~\ref{fig:tstobs} that seems consistent with the present data, and that represents
a prediction for when the sample in Fig.~\ref{fig:tstobs}, still quite small, will be hopefully enlarged:
stellar streaming always reduces $T_*$, in a way proportional to $k$ and $\alpha$,
up to an amount that can be as large as 60 percent, while flattening itself is less important
(though necessary for rotation to be important).

\begin{figure*}
\includegraphics[width=0.48\linewidth]{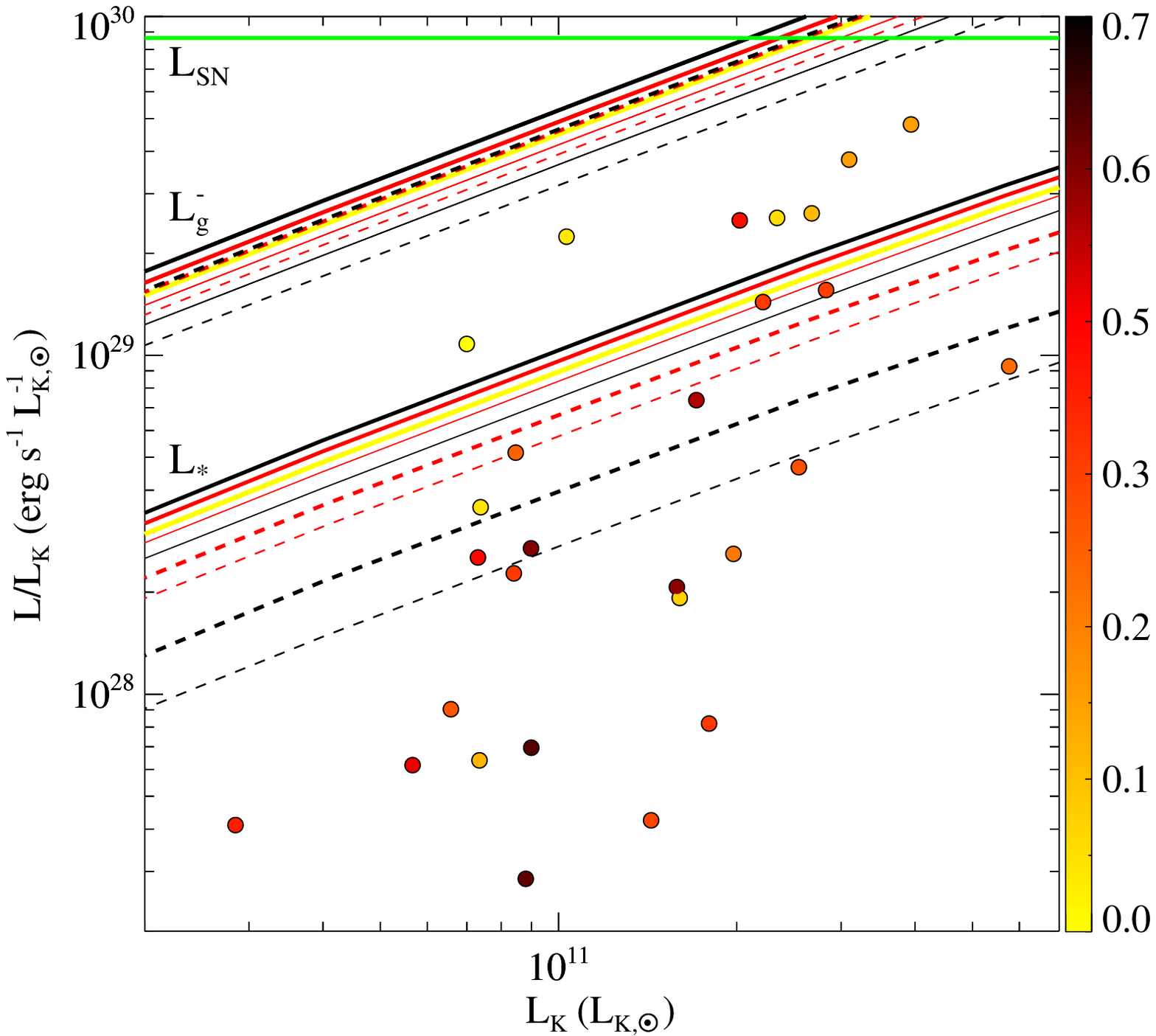}
\hskip 0.5truecm
\includegraphics[width=0.48\linewidth]{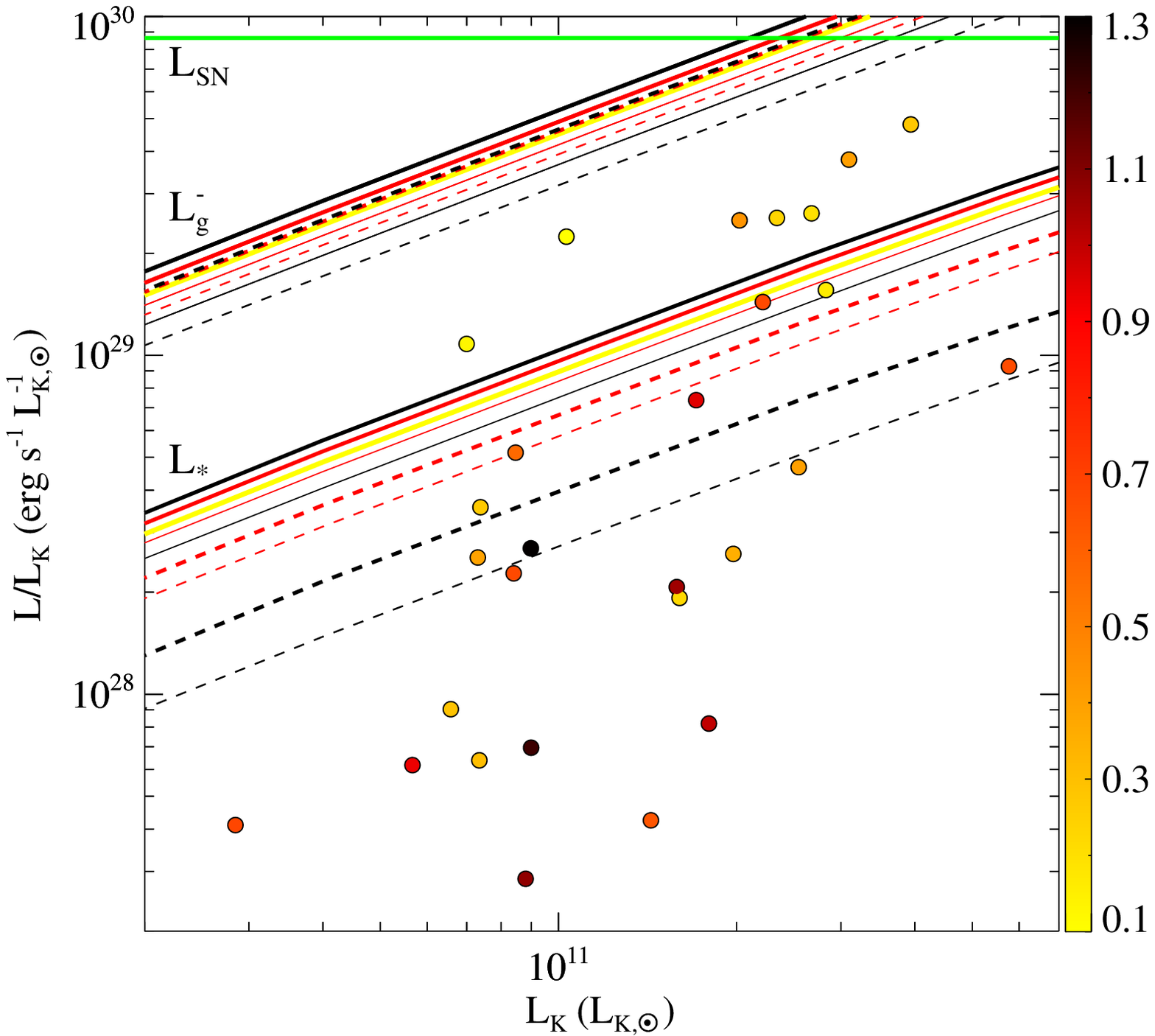}
\caption{Comparison between the run of various luminosities vs. $\Lk$, 
for all the progenitors in Tab.~\ref{tab:params} (Einasto halo case) and a selection
of their descendants (only cases of $\alpha=0,1$), and for the observed $\Lx$ for
the ETGs in the BKF sample; all quantities are normalised to $\Lk$.
The lower group of lines gives $L_*/\Lk$, the upper one gives $\Lgm/\Lk$, the horizontal line is $\Lsn/\Lk$ for $\eta=0.85$.
The yellow line refers to the progenitor ETGs, the red lines to the E4 shape, the black ones to the E7 shape;
lines are solid for $k=0$, and dashed for $k=1$ and $\alpha=1$, respectively.
Thick and thin lines refer to the FO and EO-built models, respectively.
The colour-coding for the observed ETGs indicates their ellipticity $\epsilon=1-q$ (left panel) and
the rotational support $\Vmax/\se$ (right panel), and is calibrated as for the 
previous Fig.~\ref{fig:tstobs}.}
\label{fig:tgobs}
\end{figure*}

We now move to consider possible effects from shape and rotation on the gas
content, starting from the hypothesis that it is linked to the
relative size of $\Tgm$ and $\Tinj$. Figure~\ref{fig:tgobs} shows the
observed $\Lx$ values, together with the energies
(referring to the mass of gas injected in the unit time) describing the
stellar heating $L_*$, the SNIa's heating $\Lsn$, and the requirement for
escape $\Lgm$. All quantities are normalised to
$\Lk$, and are computed using the expression for the stellar mass loss rate $\dot{M}_*$
and the temperatures defined in Sect.~\ref{sec:temp}, such that
$\Lgm=\dot{M}_*\Egm
=3k_{\mathrm B}\dot{M}_*\Tgm/( 2\mu m_{\mathrm P})$,
$L_*=3k_{\mathrm B}\dot{M}_*T_*/(2\mu m_{\mathrm P})$, and
$\Lsn=\eta E_{\mathrm{SN}}R_{\mathrm{SN}}$.
Figure~\ref{fig:tgobs} clearly shows that flatter and more rotationally supported
ETGs tend to have a lower $\Lx/\Lk$, as already known (see Sect.~\ref{sec:intro}).
As shown in Sect.~\ref{sec:eff}, however, the effect of shape or rotation on $\Lgm$ is small,
so we cannot claim an important \textit{direct} role for these two major
galactic properties on determining a lower gas content and then $\Lx$.
It has been suggested that a possible \textit{indirect} effect could come from galactic
rotation if it is effective in creating a gas disc, where gas cooling is
triggered, the temperature is lowered, and then $\Lx$ is reduced \citep{Brig}.
Other possibilities may be related to global instabilities of rotating flows, perhaps
associated with inefficient thermalisation \citep{Andrea}. Note that the X-ray emissivity is also dependent
on the gas temperature, and one could think that the lower $\Lx$ of
flat/rotating ETGs could be due to having these preferentially a lower
$\Tx$ (Fig.~\ref{fig:tstobs}). This cannot be the explanation, though, because the
emissivity in the 0.3-8 keV band decreases very mildly with
decreasing temperature, for temperatures below a value of $\sim 1$ keV.
Another explanation for a lower $\Lx/\Lk$ could be a lower stellar age in fast rotators: indeed, S13 found that
molecular gas and young stellar populations are detected only in fast
rotators across the entire ATLAS$^{\mathrm{3D}}$ sample, and a younger age
is known to be linked to a lower $\Lx$ (BKF, \citealt{O'Sullivan}).

Also, note in Fig.~\ref{fig:tgobs} how there are many ETGs with $\Lx$ lower than $L_*$: they do not even radiate $L_*$.
For them, the outflow must be very important, and must have employed almost all of $\Lsn$.
Numerical simulations (CDPR) have already shown that $\Lx$ can be even lower than $L_*$ during winds/outflows.
As $\Lgm$ increases and becomes closer to $\Lsn$, the ETGs below $L_*$ disappear.

Finally, BKF found a positive correlation between $\Lx$ and
$\Tx$, valid down to the gas-poor galaxies, that on average
have the shallowest potentials ($\Lk \lesssim2\e{11}\Lks$ and
$\se \lesssim$ 200 $\kms$) and are expected to host outflows; also in the $\Lx-\Lk$ plot, in the
large variation in $\Lx$ at any $\Lk \gtrsim 7\e{10}\Lks$,
ETGs with the lowest $\Lx$ tend to be the coldest ones. This seemed a
puzzle, if outflows (lower $\Lx$) are expected to be hotter than inflows (large $\Lx$).
P11 re-examined the $\Tx$ behavior, considering $\Tx$ values rescaled by $T_*$,
where for the latter the case of a spherical galaxy with an isotropic velocity
dispersion tensor was taken. The gas of ETGs with $\se \lesssim$ 200 $\kms$ turned out to be colder in an
absolute sense, but to have the largest values for the rescaled $\Tx$; the latter also
tend to the (rescaled) $\Tsub$, a fiducial average temperature of gas in outflow (Sect.~\ref{sec:gravtemp}),
consistent with the expectation for the gas flow status of these ETGs.
At intermediate $\se$ values, though (200 $\kms< \se <250$ $\kms$), still the rescaled $\Tx$
seemed to be lower for the gas poorest ETGs, which remained unexplained.
Can we suggest that at these intermediate $\se$ the effect of shape or rotation
is responsible for a decrease of both $\Tx$ and $\Lx$?
And then, when normalising $\Tx$ by a $T_*$ appropriate for the shape and rotational level
of the host galaxy, the result is larger than when normalising by the $T_*$ of a spherical ETG of same $\se$?
\begin{figure*}
\includegraphics[width=0.48\linewidth]{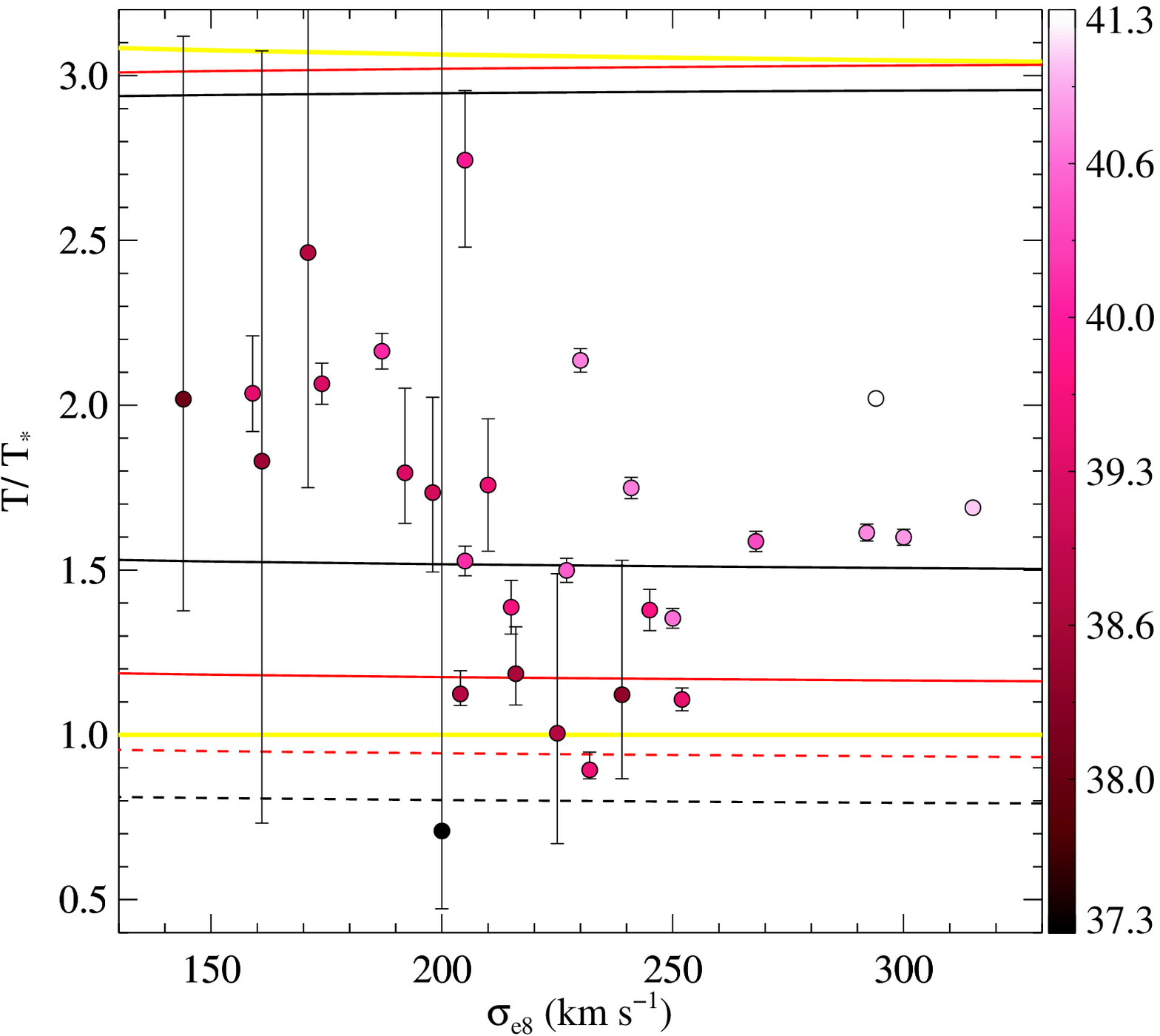}
\hskip 0.5truecm
\includegraphics[width=0.48\linewidth]{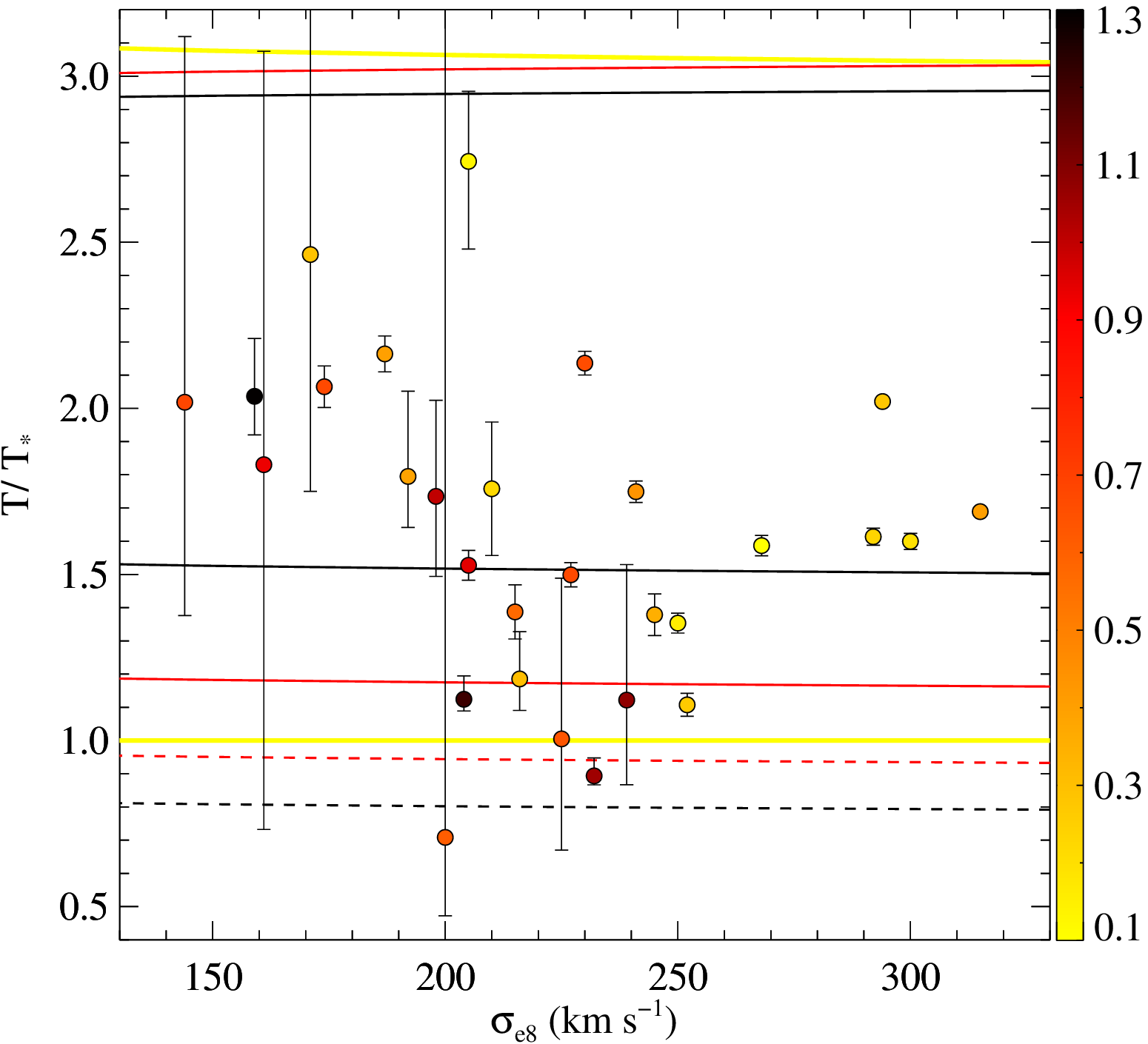}
\caption{Relative temperatures versus $\se$ for observed ETGs and models (only EO-built models
observed EO; the plot is very similar for the FO-built models). 
The colours and linetype for the models are the same as in Fig.~\ref{fig:tstobs}.
$T_*$ values (lower bundle of lines) are normalised to $T_*$ corresponding to a spherical galaxy with an isotropic velocity tensor
(given by the yellow line in Fig.~\ref{fig:tstobs}); $\Tsub$ (upper bundle of lines) and $\Tx$'s
are normalised to the $T_*$ of the corresponding shape (neglecting rotation). 
In particular, observed ETGs with $q$ in the range
[1,0.75] are normalised to $T_*$ of the E0 model, those with $q$ in the range [0.75,0.45] are normalised to
$T_*$ of the E4 model, whereas those with $q<0.45$ are normalised to $T_*$ of the E7 model.
The colour-coding for the observed ETGs indicates their log$\Lx$($\ergs$),
in the left panel, and the rotational support $\Vmax/\se$,
in the right panel, as indicated by the colour bar on the right.}
\label{fig:ratioeps}
\end{figure*}
Figure~\ref{fig:ratioeps} shows $\Tx$ values normalised by a $T_*$ appropriate
for the apparent shape of the galaxy, neglecting rotation; it confirms that at low $\se$
the gas is relatively hotter, and reaches close to $\Tsub/T_*$ (P11).
In the right panel, the colour indicates the level of rotational support $\Vmax/\se$; if rotation
is not neglected, in this plot the properly normalised $\Tx$ of rotating ETGs would
become higher (see, e.g., Fig.~\ref{fig:tstobs}).
Indeed, in the intermediate $\se$ region, some of the gas-poor ETGs with the lowest $\Tx/T_*$,
are also highly rotating, and then their position in the plot would be higher, thus at least
partially accounting for the segregation.

\section{Summary and conclusions}\label{sec:conclu}
In this paper we investigated the relationship between the temperature and luminosity of the hot
X-ray emitting haloes of ETGs and the galactic shape and rotational support. This work is an extension of
previous similar studies (CP96, \citealt{Pel97}, P11).

By solving the Jeans equations, we built a large set of axisymmetric three-component (stars, dark
matter halo, SMBH) galaxy models, representative of observed ETGs.
We varied the degree of flattening and rotational support of the stellar component, in order
to establish what is the dependence of the temperature and binding energy of the injected gas on the observed galaxy shape and 
internal kinematics. For the injected gas, we defined the equivalent temperature of stellar motions
as $T_* = T_{\sigma} + \gth T_{\mathrm{rot}}$, where the parameter $\gth$ takes into account how much
of the ordered rotation of the galaxy is eventually thermalised by the stellar mass losses.
We considered the simplified case in which the pre-existing gas velocity is
proportional to the stellar streaming velocity.
When pre-existing gas and stars rotate with the same velocity, no ordered stellar kinetic energy is thermalised
($\gth=0$); when the gas is at rest, all the
kinetic energy due to stellar streaming is thermalised ($\gth=1$). We
also defined the temperature equivalent ($\Tgm$) to the binding energy
for the injected gas. Our main results are as follows:

\begin{itemize}
\item The major effect of flattening the stellar component of a spherical ETG is
a decrease of the observed $\se$ value of its flatter counterparts of
same mass and same circularized $\re$. This decrease is proportional to the flattening level and depends
also on the viewing angle. For each shape, in velocity dispersion supported models, the decrease is larger
for the FO-view and smaller for the EO-view, and it reaches $\sim 35$ percent for the E7 shape seen FO. 
In isotropic rotators, the decrease is instead larger for the EO view.
Thus, ETGs with the same roundish appearance and the same observed $\se$ may have a significantly different mass.
This finding raises the issue of the reliability of the use
of $\se$ as a proxy for the dynamical mass of an ETG, a point particularly relevant for studies of the hot haloes
properties, that mainly depend on the galaxy mass.

\item Flatter models can be either more or less concentrated than
rounder ones of the same mass and same circularized $\re$, depending on how they are built: if
$\re$ is kept constant for a FO view, flatter models are more concentrated 
and bounded than the round counterpart; the opposite is true if $\re$ is kept constant for an EO
view. As a consequence, the effect of a pure change of shape is an
increase of $T_*$ and $\Tgm$ in the first way of flattening, and
a decrease in the second one. Overall, however, the variation in
$T_*$ for both cases is mild, within $\sim 20$ percent even for the
maximum degree of flattening (the E7 model). Similarly, $\Tgm$
gets larger by at most $\sim 20$ percent, and decreases by at most $\sim
12$ percent.

\item A more significant effect on $T_*$ can be due to the amount of
rotational support. The isotropic rotator case is investigated here ($k=1$).
If $\gth=1$, the whole stellar kinetic energy, including the streaming one, is thermalised, and $T_*$
coincides with that of the non-rotating case, for the same galaxy
shape. If $\gth<1$, $T_*$ is instead always reduced, and the
larger so the flatter is the shape. The strongest reduction is
obtained for the E7 models when $\gth=0$, and then $T_*$ can drop by $50-70$ percent
with respect to the E0 models. Thus, \textit{the presence of stellar streaming, when not thermalised,
acts always in the sense of decreasing $T_*$, and the size of this decrease depends 
on the relative motion between pre-existing gas and stars}.
Clearly, the flatter the galaxies, the stronger can be the rotational support, and, consequently its potential effect on $T_*$.
Thus the effect of rotation is dependent on the degree of flattening, and, as a minimum,
it requires a flat shape as a premise.

\item Since stellar streaming acts in the sense of making the
gas less bound due to the centrifugal support, at any fixed galaxy shape and rotational support
$\Tgm$ decreases in proportion to how the velocity field of the ISM is close to that of the stars. 
However, this decrease is lower than that produced on $T_*$: $\Tgm$ drops at most by
$\sim13$ percent (for the E7 models), between the two extreme cases of ISM at rest and
gas rotating as the stars. Note then two compensating effects from
stellar streaming when $\gth<1$: $T_*$ is lower than for a
non-rotating galaxy, but the gas is also less bound.

\item All the above trends and effects are independent of the galaxy
luminosity (mass), and dark halo shape. Only the normalization of $T_*$ and
$\Tgm$ changes, if the dark halo mass changes (e.g., it increases from
the SIS to the Einasto to the Hernquist to the NFW halo models).
\end{itemize}

The comparison of the above results with observed
$\Tx$ and $\Lx$ for the ETGs in the \textit{Chandra} sample of BKF shows that:

\begin{itemize}
\item All observed $\Tx$ are larger than $T_*$, but much closer to $T_*$ than to $\Tsn$.
$T_*$ ranges between 0.1 and 0.4 keV, for 150 $\kms\lesssim\se\lesssim300$ $\kms$
(the lower end being possibly even lower, depending on the effects of rotation),
while $\Tsn \approx 1.5$ keV (for a thermalisation parameter $\eta=0.85$),
so that the contribution from SNIa's dominates the gas injection energy ($\Tinj\sim 1.5-2$ keV),
that is then practically insensitive to changes in $T_*$ due to the galaxy shape or kinematics.
The proximity of the observed $\Tx$ values to $T_*$ indicates that $\eta$ may be lower, and/or that the
gas tends to establish itself at a temperature close to the virial temperature, in all flow phases. In any case, this
proximity provides an empirical confirmation of the relevance of a study of $T_*$ and its variations.

\item For 200 $\kms\lesssim\se\lesssim250$ $\kms$, 
$\Tgm$ becomes larger than $\Tinj$, and inflows in these galaxies
can become important. These $\se$ values could be lower
if $\eta<0.85$, and if the dark matter amount is larger than assumed here.
Galaxies with $\se\lesssim 200$ $\kms$ ($\Lk\lesssim 2\times10^{11}\Lks$), instead have a
high probability of hosting an outflow, and then a low hot gas
content, as confirmed by observations of $\Lx/\Lk$ vs. $\Lk$.
The ratio $\Lx/\Lk$ seems to be lower also for a flatter shape and larger rotation.
However, the effect of shape or rotation on $\Tgm$ is small, thus these two
major galactic properties are not expected to play an important direct
role in determining the gas content. 
Indirect effects may be more likely (as galactic rotation triggering large-scale
instabilities in the gas, or the lower age of fast rotators).
For many ETGs, $\Lx/\Lk$ is much lower than $L_*/\Lk$, indicating clearly
the presence of an outflow.

\item We find a mild indication that, at fixed $\Lk$, flatter
shapes and more rotationally supported ETGs show a
lower $\Tx$, with respect to rounder, less rotating ETGs
(similarly to what recently found by S13). This tendency can be
easily explained by the effects predicted here on $T_*$ due to
flattening and rotation. Since, for a fixed galaxy mass, a decrease of $T_*$ due
to rotation is predicted to be potentially stronger than produced by shape without rotation,
we propose that \textit{not thermalised} stellar streaming is a more efficient cause of
the (possibly) lower $\Tx$.

\item Extending a P11 result, we find that, when rescaled by a $T_*$
value representative of a fully velocity dispersion supported ETG of the corresponding shape,
the $\Tx/T_*$ values of outflows (at $\se\lesssim 200$ $\kms$) are larger than those of
inflows (at $\se\gtrsim 250$ $\kms$). At 200 $\kms\lesssim\se\lesssim250$ $\kms$,
where the $\Tx/T_*$ values seem to be lower for the gas poorest ETGs, part of the trend
could be explained by a few of these ETGs being highly rotating.
Note that the observed galactic rotation depends on the viewing angle at which these low
$\Tx$ and low $\Lx$ ETGs are seen, thus it is difficult to find a clear indication
of a systematic trend in the data.
\end{itemize}

As a necessary parallel investigation, we are going to perform 2D hydrodynamical simulations
to explore the actual behaviour of the ISM for the models here studied \citep{Andrea}.
The goal will be to establish, as a function of galactic structure and kinematics, the
presence of major hydrodynamical instabilities, the efficiency of thermalisation of
stellar streaming motions, and their impact on $\Tx$ and $\Lx$.

\section*{Acknowledgments}
We thank M. Cappellari and D.-W. Kim for useful discussions.
L.C. and S.P. were supported by the MIUR grants PRIN 2008 and PRIN 2010-2011, project `The 
Chemical and Dynamical Evolution of the Milky Way and Local Group 
Galaxies', prot. 2010LY5N2T. This material is based upon work
of L.C. and S.P. supported in part by the National Science Foundation under Grant No.
1066293 and the hospitality of the Aspen Center for Physics.

\bibliographystyle{mn2e}
\bibliography{references}{}

\appendix
\section{The fluid equations in the presence of source terms}
\label{sec:flueqs}
The equations of fluid dynamics in the presence of different sources of mass, momentum and
energy, under the simplifying assumption of isotropy of the mass losses, can be written \citep{Derc} as
\begin{gather}
 \dfrac{D\rho}{Dt} + \rho(\nabla \cdot \Bld{u}) = \mathscr{M}, \\
 \rho \dfrac{D\Bld{u}}{Dt} =- \rho \nabla\Phi - \nabla p + \mathscr{M}
(\Bld{v}-\Bld{u}), 
\label{eq:1fe}\\
\begin{split}
\dfrac{DE}{Dt} + (E + p)\nabla \cdot \Bld{u}=
&\, \Sigma\mathscr{M}_i\left[e_i +\dfrac{u_{\mathrm{s},\,i}^2}{2}\right] -\mathscr{L}\\
&+\dfrac{\mathscr{M}}{2}\left[\Arrowvert\Bld{u}-\Bld{v}\Arrowvert^2 +
 \text{Tr}(\Bld{\sigma}^2)\right].
 \label{eq:3fe}
\end{split}
\end{gather}
Here $\mathscr{M}(\mathbf{x},t)=\Sigma\mathscr{M}_i$ is the total mass return per unit time and
volume, due to different sources (e.g., stellar winds and SNIa events) associated with a
stellar population with streaming velocity $\Bld{v}(\mathbf{x},t)$ and velocity dispersion $\Bld{\sigma}^2(\mathbf{x},t)$.
$e_i(\mathbf{x},t)$ is the internal energy return per unit mass and time of the $i$-th source field,
and $u_{\mathrm{s},\,i}(\mathbf{x},t)$ is the modulus of the relative velocity of the material injected by the $i$-th source field 
with respect to the source star. Finally, $\mathscr{L}$ are the bolometric radiative losses per unit time and volume.

For example, in applications as the one in this paper, $\mathscr{M}$ is represented by the sum
of stellar winds and SNIa explosions ejecta. Often, $e_{\mathrm{wind}}$ and $u_{\mathrm{s,wind}}$
are neglected, being significantly smaller than the contribution of $\text{Tr}(\Bld{\sigma}^2)$, while
the opposite holds for SNIa events, where $\mathscr{M}_{\mathrm{SN}}\text{Tr}(\Bld{\sigma}^2)$ is
negligible with respect to the energy injection due to the SNIa explosions.
In our case, $\gth$ in Eq.~(\ref{eq:gth}) derives from the full mass injection term in Eq.~(\ref{eq:3fe}),
where both contributions are taken into account.

\section[]{The code} \label{sec:code}
All the relevant dynamical properties of the models are computed using a
numerical code built on purpose \citep{ProcMio,tesi}. Starting from an axisymmetric density
distribution the code computes in terms of elliptic integrals the associated gravitational potential and
vertical and radial forces. The code then solves the Jeans equations in cylindrical
coordinates $(R,z,\varphi)$. For an axisymmetric density distribution $\rho_*(R,z)$ supported by a
two-integral phase-space distribution function (DF), the Jeans equations write
\beq
\dfrac{\partial\rho_*\sigma^2}{\partial
z}=-\rho_*\dfrac{\partial\Phi_{\mathrm{tot}}}{\partial z},
\label{eq:J1}
\eeq
and
\beq
\dfrac{\partial\rho_*\sigma^2}{\partial
R}+\rho_*\dfrac{\sigma^2-\overline{\vphi^2}}{R}=-\rho_*
\dfrac{\partial\Phi_{\mathrm{tot}}}{\partial R},
\label{eq:J2}
\eeq
(e.g. \citealt{BinTre}), where $\Phi_{\mathrm{tot}}$ is the sum of the gravitational potentials of all the
components (e.g. stars, dark halo, black hole).
As well known, for a two-integral DF (1) the velocity dispersion tensor
is diagonal and aligned with the coordinate system; (2) the
radial and vertical velocity dispersions are equal, i.e.
$\sigma_R=\sigma_z\equiv\sigma$; (3) the only non-zero streaming motion is in
the azimuthal direction.

In order to control the amount of ordered azimuthal velocity $\overline{\vphi}$, 
we adopted the $k$-decomposition introduced by \cite{Satoh}
\beq
\overline{\vphi}^2=k^2(\overline{\vphi^2}-\sigma^2),
\eeq
and then it follows
\beq
\sigma_{\varphi}^2\equiv\overline{\vphi^2}-\overline{\vphi}
^2=\sigma^2+(1-k^2)(\overline{\vphi^2}-\sigma^2),
\eeq
where $0\leqslant k \leqslant 1$. The case $k=1$ corresponds to the isotropic
rotator, while for $k=0$ no net rotation is present and all the flattening is due
to the azimuthal velocity dispersion $\sigma_{\varphi}$.
In principle, $k$ can be a function of $(R,z)$, and so more complicated (realistic)
velocity fields can be realized (CP96; see also \citealt{ProcAnd}).
The code then projects all the relevant kinematical fields, together with the stellar density.

The projections along a general l.o.s. of the stellar density $\rho_*$,
streaming velocity $\Bld{v}$ and velocity dispersion tensor $\Bld{\sigma}^2$ are
\beq
\Sigma_*=\int_{-\infty}^{+\infty}\rho_*dl,
\eeq
\beq
\Sigma_*v_{\mathrm{los}}=\int_{-\infty}^{+\infty} \rho_*\avg{\Bld{v},\Bld{n}} dl,
\eeq
\beq
\Sigma_*\sigma^2_{\mathrm P}=
\int_{-\infty}^{+\infty}\rho_*\avg{\Bld{\sigma}^2\Bld{n},\Bld{n}} dl,
\eeq
respectively, where $\avg{,}$ is the scalar product, $\Bld{n}$ is the l.o.s. direction
and $l$ is the integration path along $\Bld{n}$.
Note that if a rotational support is present, then $\sigma^2_{\mathrm P}$ is not the
l.o.s. (i.e. the observed) velocity dispersion $\sigma_{\mathrm{los}}^2$, given by
\beq
\sigma^2_{\mathrm{los}}=\sigma_{\mathrm P}^2+V_{\mathrm P}^2-v_{\mathrm{los}}^2,
\eeq
where $V_{\mathrm P}^2$ is the projection of $\avg{\Bld{n},\Bld{v}}^2$ (CP96).
In particular, the face-on projections are 
\beq
\Sigma_*=2\int_0^\infty \rho_*dz,
\eeq
\beq
\Sigma_*\sigma^2_{\mathrm P}=2\int_0^\infty \rho_*\sigma^2 dz,
\eeq
with $\sigma^2_{\mathrm{los}}=\sigma^2_{\mathrm P}$.
The edge-on projections are instead
\beq
\Sigma_*=2\int_R^\infty \dfrac{\rho_*\tilde{R}\,d\tilde{R}}{\sqrt{\tilde{R}^2-R^2}},
\eeq
\beq
\Sigma_*v_{\mathrm{los}}=2R\int_R^\infty\dfrac{\rho_*\overline{\vphi}\,d\tilde{R}}{\sqrt{\tilde{R}^2-R^2}},
\eeq
\beq
\Sigma_*\sigma_{\mathrm P}^2=2\int_{R}^{\infty}
\left[(\tilde{R}^2-R^2)\sigma^2+
R^2\sigma^2_{\varphi}\right]\dfrac{\rho_*
\,d\tilde{R}}{\tilde{R}\sqrt{\tilde{R}^2-R^2}},
\eeq
\beq
\Sigma_*V_{\mathrm P}^2=2R^2\int_R^\infty \dfrac{\rho_*\overline{\vphi}^2
\,d\tilde{R}}{\tilde{R}\sqrt{\tilde{R}^2-R^2}},
\eeq
where all the integrations are performed at fixed $z$.

After the projection, the code calculates the corresponding circularized effective radius
$\re$. In practice, the stellar projected density is integrated on the isodensity curves,
and $\re^2=q_{\mathrm{los}}a_{\mathrm e}^2$, where $a_{\mathrm e}$ and $q_{\mathrm{los}}$
are the semi-major axis and l.o.s. axial ratio of the ellipse of half projected luminosity, respectively.
In general, $q_{\mathrm{los}}$ is a function of the l.o.s. inclination \citep{LanzCioz}: for
a face-on projection $q_{\mathrm{los}}=1$, while for the edge-on case $q_{\mathrm{los}}=q$
(see Sect.~\ref{sec:mods_star}).
Then, the corresponding luminosity averaged aperture velocity dispersion
\beq
\se^2\equiv\dfrac{\int_0^{\re/8}
\Sigma_*\sigma_{\mathrm{los}}^2RdR}{\int_0^{\re/8} \Sigma_*RdR},
\label{eq:se8}
\eeq
is computed within a circular aperture of $\re/8$.

For comparison with other works, we followed also the approach of the ATLAS$^{\mathrm{3D}}$
project \citep{CappATLAS3D,CappXV}, calculating the quantity
\beq
\Vrms^2=\sigma_{\mathrm P}^2+V_{\mathrm P}^2=\sigma^2_{\mathrm{los}}+v_{\mathrm{los}}^2,
\label{eq:vrms}
\eeq
and its corresponding luminosity averaged mean within $\re/8$, according to a definition
analogous to Eq.~(\ref{eq:se8}).
Finally, for a given model, the code evaluates all the volume integrals in Sect.~\ref{sec:temp},
by using a standard finite-difference scheme.

\label{lastpage}

\end{document}